%% file: Dust_II.tex
\documentclass[a4paper,usenatbib,times]{mn2e}
\usepackage{natbib,graphicx,amsmath,amsfonts,amssymb,times,txfonts}

\def\vg{\textbf{v}_{\mathrm{g}}}
\def\vd{\textbf{v}_{\mathrm{d}}}
\def\rhog{\rho_{\mathrm{g}}}
\def\hrhog{\hat{\rho}_{\mathrm{g}}}
\def\rhod{\rho_{\mathrm{d}}}
\def\hrhod{\hat{\rho}_{\mathrm{d}}}

\def\va{\textbf{v}_{a}}

\def\vb{\textbf{v}_{b}}

\def\Pga{P_a}

\def\hrhoga{\hat{\rho}_{a}}

\def\hrhoda{\hat{\rho}_{i}}
\def\thetaa{\theta_{a}}
\def\ma{m_{a}}

\def\doha{\left(h_{a}  \right)}

\def\omga{\Omega_{a} }

\def\Rdak{R_{\mathrm{d},ak}}

\def\Pgb{P_{b}}
\def\hb{h_{b}}
\def\hrhogb{\hat{\rho}_{b}}

\def\hrhodb{\hat{\rho}_{{\mathrm{d},b}} }
\def\thetab{\theta_{b}}
\def\mb{m_{b}}

\def\dohb{\left(h_{b}  \right)}

\def\omgb{\Omega_{b} }
\def\omdb{\Omega_{\mathrm{d},b} }

\def\mj{m_{j}}

\def\mk{m_{k}}

\def\mc{m_{c}}

\def\deltav{\Delta v}
\def\tdeltav{\Delta \textbf{v}}
\def\cs{c_{\mathrm{s}}}

\def\Rd{R_{\mathrm{d}}}
\def\cd{C_{\mathrm{D}}}

\def\kak{K_{ak}}
\def\vak{\textbf{v}_{ak}}
\def\hrak{\hat{\textbf{r}}_{ak}}

\def\hrbi{\hat{\textbf{r}}_{bi}}

\def\hraj{\hat{\textbf{r}}_{aj}}

\def\mi{m_{i}}

\def\Pg{P_{\mathrm{g}}}

\def\vgi{v_{\mathrm{g},i}}
\def\vdi{v_{\mathrm{d},i}}

\def\dst{\displaystyle}

\newcommand{\kn}[2]{W_{\!\!\ #1\!\! \ #2}}
\newcommand{\gkn}[3]{\nabla_{\!\! \ #1}W_{\!\! \ #2\!\! \ #3}}

\title[Dusty gas with SPH --- II.]{Dusty gas with SPH --- II. Implicit timestepping and astrophysical drag regimes}

\author[Laibe \& Price]{Guillaume Laibe, Daniel J. Price \\
Centre for Stellar and Planetary Astrophysics and School of Mathematical Sciences, Monash University, Clayton, Vic 3800, Australia
}
\pagerange{\pageref{firstpage}--\pageref{lastpage}} \pubyear{2011}

\begin{document}
\include{journaux}

\label{firstpage}
\bibliographystyle{mn2e}
\maketitle

\begin{abstract}
In a companion paper  \citep{LP11b}, we have presented an algorithm for simulating two-fluid gas and dust mixtures in Smoothed Particle Hydrodynamics (SPH). In this paper, we develop an implicit timestepping method that preserves the exact conservation of the both linear and angular momentum in the underlying SPH algorithm, but unlike previous schemes, allows the iterations to converge to arbitrary accuracy and is suited to the treatment of non-linear drag regimes. The algorithm presented in Paper~I is also extended to deal with realistic astrophysical drag regimes, including both linear and non-linear Epstein and Stokes drag. The scheme is benchmarked against the test suite presented in Paper~I, including i) the analytic solutions of the \textsc{dustybox} problem and ii) solutions of the \textsc{dustywave}, \textsc{dustyshock}, \textsc{dustysedov} and \textsc{dustydisc} obtained with explicit timestepping. We find that the implicit method is 1--10 times faster than the explicit temporal integration when the ratio $r$ between the the timestep and the drag stopping time is $1 \lesssim r \lesssim 1000$. 

\end{abstract}


%----------------------------------------------------------------------------------------------------------------
\section{Introduction}

Dust in cold astrophysical systems spans a huge range of sizes from sub-micron sized grains in the interstellar medium to kilometre sized planetesimals involved in planet formation. Moreover, the ratio of dust to gas as well as the density and temperature of the gaseous environment in which dust is embedded can also vary strongly. Handling this full range of physical parameters presents a challenge to numerical schemes designed to simulate dusty gas in astrophysics. The main challenges are i) that at high drag (e.g. small grains), the small timestep required means that purely explicit timestepping methods become prohibitive and ii) that a wide range of physical drag prescriptions, including non-linear drag regimes, need to be handled by the code.

Two main prescriptions for drag between gas and solid particles are applicable to astrophysics: The Epstein regime --- where the gas surrounding a grain can be treated as a dilute medium --- and the Stokes regime --- where the grains can be treated as solid bodies surrounded by a fluid ---(see e.g. \citealt{Baines1965,Stepinski1996}). The dependance of the drag term on local parameters of the gas (density, temperature) and the dust (typical grain size, mass) differ between the two regimes, in turn implying very different dynamics for the dust grains. For example, in a protoplanetary disc, both of these regimes may be applicable in different regions of the disc.

In a companion paper (\citealt{LP11b}, hereafter Paper~I), we have developed a new algorithm for treating two-fluid gas-dust astrophysical mixtures in Smoothed Particle Hydrodynamics (SPH). Benchmarking of the method demonstrated that the algorithm gives accurate solutions on a range of test problems relevant to astrophysics and substantially improve previous algorithms \citep{Monaghan1995}. However, in Paper~I, we used only a simple explicit time stepping and considered only linear drag regimes with a constant drag coefficient. In this paper, we present an implicit timestepping method that can be applied to both linear and non-linear drag regimes, which is both more accurate and more general than the scheme proposed by \citet{Monaghan1997}. We also discuss the SPH implementation of both the Epstein and the Stokes regimes in their full generality.

The paper is organised as follows: The equations of motion and the characteristics of the different astrophysical drag regimes for gas and dust mixtures are given in Sec.~\ref{sec:two_fluids}. We summarise the SPH formalism used for integrating these equations (derived in detail in Paper~I) in Sec.~\ref{sec:two_fluids_SPH} and extend it to deal with the drag regimes encountered in astrophysics. Particular attention is paid in Sec.~\ref{sec:timestepping} to improving the \citet{Monaghan1997} implicit timestepping scheme, including its generalisation for non-linear drag regimes. Finally, the algorithm for non-linear drag regimes is tested against the analytic solutions of the \textsc{dustybox} problem, as well as the \textsc{dustywave}, \textsc{dustysedov}, \textsc{dustyshock} and \textsc{dustydisc}  tests, in Sec.~\ref{sec:tests}.
%=======================================================================================================
\section{Gas and dust evolution in astrophysical systems}
\label{sec:two_fluids}

\subsection{Evolution equations}

The equations describing the evolution of astrophysical gas and dust mixtures, where dust is treated as a pressureless, inviscid, continuous fluid have been described in detail in Paper~I. The equations in the continuum limit are given by:
\begin{eqnarray}
\frac{\partial \hrhog}{\partial t} + \nabla . \left ( \hrhog \vg \right) & = & 0 \label{eq:mass_gas},\\
\frac{\partial \hrhod}{\partial t} + \nabla . \left ( \hrhod \vd \right) & = & 0 \label{eq:mass_dust},\\
\hrhog \left( \frac{\partial \vg}{\partial t} + \vg . \nabla \vg \right)  & = & - \theta\phantom{.}\nabla P_{\rm g} + \hrhog \textbf{f} -  F_{\rm drag}^{\rm V} \label{eq:momentum_gas},\\
\hrhod \left( \frac{\partial \vd}{\partial t} + \vd . \nabla \vd \right) & = &   - \left(1 -\theta \right) \nabla P_{\rm g} + \hrhod \textbf{f} +  F_{\rm drag}^{\rm V} \label{eq:momentum_dust},\\
\frac{{\rm d}u_{{\rm g}}}{{\rm d}t} & = & -\frac{\Pg}{\hrhog} \left[ \theta \nabla\cdot{\bf v}_{\rm g} + (1 - \theta) \nabla\cdot{\bf v}_{\rm d} \right] +  \Lambda_{\rm drag} \label{eq:int_en}.
\end{eqnarray}
where the subscripts ${\rm g}$ and ${\rm d}$ refer to the gas and dust, respectively such that $P_{\rm g}$ is the gas pressure, $\vg$ and $\vd$ are the fluid velocities and $u$ is the specific internal energy of gas. The volume densities of gas and dust ($\hrhog$ and $\hrhod$, respectively) are related to the corresponding intrinsic densities ($\rho_{\rm{g}}$ and $\rho_{\rm{d}}$, respectively) according to
\begin{eqnarray}
\hrhod & = & (1 - \theta) \rhod, \\
\hrhog & = & \theta \rhog,
\end{eqnarray}
where $\theta$ is the volume filling fraction of the dust. Finally, the drag force and heating terms are given by:
\begin{equation}
 F_{\rm drag}^{\rm V}  =  K  (\vg - \vd) , \label{eq:fdrag}
 \end{equation}
 and
 \begin{equation}
  \Lambda_{\rm drag}  = K (\vg - \vd)^{2} \label{eq:endrag}.
\end{equation}
 The drag coefficient K has dimensions of mass per unit volume per unit time and is generally a function of the relative velocity between the two fluids $\Delta v \equiv \vert \vg - \vd \vert$, implying a non-linear drag regime with respect to the differential velocity between the gas and the dust. In most of the astrophysical systems, the dust is diluted enough into the gas so that the gas filling fraction is $\theta \simeq 1$ to a very good level of approximation, such that the dust buoyancy term $\left(1 -\theta \right) \nabla P_{\rm g}$ is negligible.

\subsection{Astrophysical drag regimes}

Microscopic collisions of gas molecules on a single dust grain result in a net exchange of momentum which is equivalent to a drag force $\bf{F}_{\rm{drag}}$ between the two phases. Two limiting cases occur when comparing the typical geometrical size $s$ of a dust grain to the mean free path $\lambda_{\rm{g}}$ of the gas. 

When the typical grain size is negligible compared to the collisional mean free path of the gas particles ($s \ll \lambda_{\rm{g}}$), the grains are surrounded by a dilute gas phase and may be treated using the Epstein drag prescription. In this limit, the analytic expression of the resulting drag force has been derived, assuming spherical, compact grains with homogeneous composition, for both specular and diffuse reflections on the grain surface (see \citealt{Baines1965} for the complete derivation). These expressions have been widely used in astrophysical studies (see e.g. \citealt{ChiangYoudin2010} for references), sometimes incorrectly where the grains are known to be porous and have fractal structures \citep{Blumwrum2008} (thus breaking the assumptions of the Epstein prescription).

For grain sizes larger than the collisional mean free path ($s \gg \lambda_{\rm{g}}$), grains experience a local differential velocity with respect to a uniform viscous flow and should be treated using the Stokes drag prescription \citep{FanZhu}. In this case, the momentum is diffused by viscosity into the fluid, which implies that the drag expression strongly depends on the \textit{local} Reynolds number defined according to
\begin{equation}
\Rd= \frac{2s  \left| \textbf{v}_{\mathrm{d}} - \textbf{v}_{\mathrm{g}}\right|}{\nu} ,
\label{eq:def_Rd}
\end{equation}
where $\nu$ is the kinematic viscosity of the gas. Analytic expressions for the drag force can be derived at small Reynolds numbers. At higher Reynolds number, the drag law is inferred from experiments. Rigorously, additional contributions to the drag should arise from the grain acceleration (carried mass and Basset contribution), grain rotation (Magnus) and in the presence of strong local shear, pressure and temperature gradients (e.g.  \citealt{FanZhu}). These corrections are negligible in nearly all astrophysical contexts.

No current analytic theory describes how both the gas and the dust fluid exchange momentum in the intermediate regime (i.e. $s \simeq \lambda_{\rm{g}}$). Generally, an asymptotic continuous interpolation between the two limiting Epstein and Stokes regimes is used. \citet{Stepinski1996} suggest adopting $s = \frac{4}{9} \lambda_{\rm{g}}$ as a means of obtaining a smooth transition. It should be noted that although this approach is convenient,  there is no clear measure of the physical accuracy of this assumption.

\subsubsection{Epstein regime for dilute media}

In a dilute medium ($\lambda_\mathrm{g}>{4s}/{9}$), grains are small enough not to disturb the Maxwellian distribution of the gas velocity. Assuming grains are spherical, that the mass of a gas molecule is negligible compared to the mass of a dust grain and that the reflection of gas particles from collisions  with dust grains are specular, the expression of the drag force \textit{on a single grain} $\bf{F}_{\rm{drag}}$ (which differs from the \textit{volume} force $F_{\rm drag}^{\rm V}$ by a factor $\hrhod/m_{\rm d}$, see Paper~I) for the Epstein regime is given by 
\begin{align}
\mathbf{F}_{\rm{drag}} =  \dst - 2 \pi s^{2}\rhog \deltav ^{2} & \left[ \frac{1}{2\sqrt{\pi}} \left\lbrace \left( \frac{1}{\psi} + \frac{1}{2 \psi^3} \right)e^{-\psi^{2}} + \right. \right.  \label{eq:Epstein_specular} \\  \nonumber
& \left. \left.  \left(  1 + \frac{1}{\psi^{2}} - \frac{1}{4 \psi^{4}}  \right) \sqrt{\pi} \, \mathrm{erf} \left( \psi \right)  \right\rbrace \right]  \textbf{x} ,
\end{align} 
where $s$ corresponds to the grain radius and $m$, $\rhog$, $T$ denote the mass of the gas molecules, the intrinsic gas density and the local temperature of the mixture (the gas and the dust are supposed to have the same temperature). The thermal sound speed of the gas is thus $\cs = \sqrt{\gamma k_{\mathrm{B}} T/m }$ and the mean thermal velocity of the gas, $ \cs \sqrt{8/ \pi \gamma}$. The dimensionless quantity $\psi$ is defined according to
\begin{equation}
\psi  \equiv \dst\sqrt{ \frac{\gamma}{2}} \frac{\deltav}{\cs} ,
\label{eq:def_psi}
\end{equation}
where $\tdeltav = \vd - \vg = \deltav$ \textbf{x} is the differential velocity (\textbf{x} being a unit vector). However, depending on the characteristics of the problem (i.e. low or high Mach numbers, or both), simpler and computationally less expensive approximations may be used. For $\psi \ll 1$, i.e. low Mach numbers, Eq.~\ref{eq:Epstein_specular} can be expanded to third order in $\psi$, giving
\begin{equation}
\mathbf{F}_{\rm{drag}} = \dst -\frac{4\pi}{3} \rhog s^2 \sqrt{\frac{8}{\pi \gamma}} \cs \deltav \left[1 +  \frac{\psi^{2}}{5} + \mathcal{O}\left(\psi^{4} \right)    \right] \textbf{x}, 
\label{eq:Epstein_third}
\end{equation}
which is usually simplified to its linear term,
\begin{equation}
\mathbf{F}_{\rm{drag}} = -\frac{4\pi}{3} \rhog s^2 \sqrt{\frac{8}{\pi \gamma}} \cs \Delta \mathbf{v} .
\label{eq:Epstein_linear}
\end{equation}
For  $\psi \gg 1$, i.e. high Mach numbers, the Taylor expansion in $1/\psi$ of Eq.~\ref{eq:Epstein_specular} gives
\begin{equation}
\mathbf{F}_{\rm{drag}} = \dst - \left[ \pi \rhog s^{2} \deltav ^{2}\left(1 + \frac{1}{\psi^{2}} - \frac{1}{4\psi^{4}} \right) + \mathcal{O}\left( e^{-\psi^{2}} \right)  \right]  \textbf{x}, 
\label{eq:Epstein_stuff}
\end{equation}
which is usually reduced to its quadratic term,
\begin{equation}
\mathbf{F}_{\rm{drag}} = -  \pi \rhog s^{2} \deltav \Delta \mathbf{v} .
\label{eq:Epstein_quad}
\end{equation}
A convenient way to handle Epstein drag at both low and high Mach numbers is to use an interpolation between the two asymptotic regimes given by Eqs.~\ref{eq:Epstein_linear} and \ref{eq:Epstein_quad} as derived in \citet{Kwok1975} (cf. \citealt{PM2006}), giving
\begin{equation}
\mathbf{F}_{\rm{drag}} = -\frac{4\pi}{3} \rhog s^2 \sqrt{\frac{8}{\pi \gamma}} \cs \sqrt{1 + \frac{9 \pi}{128} \frac{\Delta v ^{2}}{c_{\rm{s}}^{2}} } \Delta \mathbf{v} .
\label{eq:Epstein_mixed}
\end{equation}
The deviation of Eq.~\ref{eq:Epstein_mixed} from the full expression (Eq.~\ref{eq:Epstein_specular}) is $\lesssim 1\%$ \citep{Kwok1975}. Thus, in general, we adopt Eq.~\ref{eq:Epstein_mixed} for the Epstein regime. We compare the differences between the various Epstein expressions in Sec.~\ref{sec:tests}.

\subsubsection{Stokes regime for dense media}
In a dense medium ($\lambda_\mathrm{g}>{4s}/{9}$), grains should be treated with the Stokes drag regime, for which the expression of the drag force $\bf{F}_{\rm{drag}}$ is:
\begin{equation}
\mathbf{F}_{\rm{drag}} = -\frac{1}{2}\cd \pi s^{2} \rhog \deltav \tdeltav ,
\label{eq:BBO_generalised}
\end{equation}
where the coefficient $\cd$ is a piecewise function of the local Reynolds number:
\begin{equation}
\cd =
\begin{cases}
24 \Rd^{-1} , &  \Rd < 1 ; \\
24 \Rd^{-0.6} ,& 1 < \Rd < 800 ;\\
0.44, & 800 < \Rd,
\end{cases}
\label{eq:cd_coefficients}
\end{equation}
where $\Rd$ is defined in Eq.~\ref{eq:def_Rd}. Equation~\ref{eq:cd_coefficients} indicates that at small Reynolds numbers ($\Rd < 1$), the drag force is linear with respect to the local differential velocity between the grain and the gas. The relation transitions to a power-law regime ($\mathbf{F}_{\rm{drag}} \propto \deltav ^{0.4} \tdeltav$) at intermediate Reynolds numbers ($1 < \Rd < 800$) and becomes quadratic at large Reynolds numbers ($\Rd > 800$). When the local concentration of dust grains becomes very large (i.e., average distance between the particles comparable to the grain size), the coefficient $\cd$ should also depend on the local concentration of particles. However, this extreme situation is not encountered in astrophysical situations. 

Assuming gas molecules interact as hard spheres, the dynamic viscosity of the gas can be computed according to \citep{ChapmanCowling}:
\begin{equation}
\mu = \frac{5m}{64 \sigma_{\mathrm{s}}}\sqrt{\frac{\pi}{\gamma}} \cs ,
\label{eq:viscosity_hard}
\end{equation}
where $m = 2m_{\mathrm{H}}$ and  $\sigma_{\mathrm{s}}$ is the geometric cross section of the molecule ($\sigma_{\mathrm{s}} = 2.367 \times10^{-15}$ cm$^{2}$ for H$_{2}$). The gas mean free path $\lambda_{\mathrm{g}}$ and the kinematic viscosity $\nu$ of the gas are deduced from $\mu$ using
\begin{equation}
\lambda_{\mathrm{g}}  = \dst  \sqrt{\frac{ \pi \gamma }{2 }} \frac{ \mu}{ \rhog \cs} ,
\label{eq:gas_mfp}
\end{equation}
and
\begin{equation}
\nu  =  \dst\frac{\mu}{\rhog} .
\label{eq:gas_nu}
\end{equation}
%=======================================================================================================
\section{Asytrophysical dust and gas mixtures in SPH}
\label{sec:two_fluids_SPH}

\subsection{SPH evolution equations}

The SPH version of the continuity equations Eqs.~\ref{eq:mass_gas} -- \ref{eq:mass_dust} are given by the density summations for both the gas and the dust phase, computed according to:
\begin{eqnarray}
\hrhoga = \sum_{b} m_{b}  W_{ab} (h_{a}); &\hspace{1cm} & h_{a} = \eta \left(\frac{\ma}{\hrhoga} \right )^{1/\nu}, \label{eq:h_density_gas} \\
\hrhoda = \sum_{j} m_{j}  W_{ij} (h_{i});  &\hspace{1cm}& h_{i} = \eta \left(\frac{m_{j}}{\hrhoda} \right )^{1/\nu},
\label{eq:h_density_dust}
\end{eqnarray}
where as in Paper~I, the indices $a,b,c$ refer to quantities computed on gas particles and $i,j,k$ refer to quantities computed on dust particles. The volume filling fraction $\theta$, is defined on a \emph{gas} particle, $a$, according to
\begin{equation}
\thetaa = 1 - \frac{ \hat{\rho}_{{\rm d},a} }{ \rho_{\rm d} },
\label{eq:SPH_theta}
\end{equation}
where $ \hat{\rho}_{{\rm d},a}$ is the density of \emph{dust} at the \emph{gas} particle location, calculated using
\begin{equation}
\hat{\rho}_{{\rm d},a} = \sum^{N_{neigh, dust}}_{j=1} m_{j} W_{aj} (h_{a}),
\label{eq:hdensitydustatgas}
\end{equation}
where $h_{a}$ is the smoothing length of the \emph{gas} particle computed using gas neighbours. The local density of dust at the gas location can thus be zero (giving $\theta = 1$) if no dust particles are found within the kernel radius computed with the gas smoothing length. Importantly, as $\hat{\rho}$ and $h$ are mutually dependent, they have to be simultaneously calculated for each type of particle, e.g. by the iterative procedure described in \citet{pm07}.

The SPH equations of motion for the gas and the dust particles, corresponding to the SPH translation of Eqs.~\ref{eq:momentum_gas} and \ref{eq:momentum_dust}, are given by
\begin{align}
 \frac{\mathrm{d} \va}{\mathrm{d}t}  = &  - \sum_{b} \mb \left[ \frac{\Pga \tilde{\theta}_{a}}{\omga\hrhoga^{2}} \gkn{a}{a}{b} \doha + \frac{\Pgb \tilde{\theta}_{b}}{\omgb\hrhogb^{2}} \gkn{a}{a}{b} \dohb \right]  \label{eq:mom_SPH_gas} \nonumber \\ 
 & - \sum_{j} \mj \frac{\Pga \left(1 - \thetaa \right )}{\hrhoga  \hat{\rho}_{{\rm d},a}} \gkn{a}{a}{j} \doha \nonumber \\
 & + \nu \sum_{j} m_{j} \frac{K_{aj}}{\hat{\rho}_{a} \hat{\rho}_{j}} \left({\bf v}_{aj} \cdot \hraj \right) \hraj D_{aj} (h_{a}) ,
\end{align}
for an SPH gas particle and 
\begin{align}
\frac{\mathrm{d} {\bf v}_{i}}{\mathrm{d}t} = & \sum_{b} \mb \frac{ \Pgb \left(1 - \thetab \right) }{\hrhogb \hrhodb } \gkn{i}{b}{i} \dohb \label{eq:mom_SPH_dust} \\ \nonumber 
& - \nu \sum_{b} \mb \frac{K_{bi}}{\hat{\rho}_{b} \hat{\rho}_{i}} \left({\bf v}_{bi} \cdot \hrbi \right) \hrbi D_{ib}  (h_{i}) ,
\end{align}
for an SPH dust particle. $\Omega$ is the usual variable smoothing length term
\begin{equation}
\Omega_{b} \equiv 1 -  \frac{\partial \hb }{\partial \hrhogb} \sum_{c} \mc \frac{\partial \kn{b}{c}  \dohb }{\partial \hb} .
\end{equation}
It should be noted that $\Omega_{\rm{d}}$ is computed only using dust particle neighbours according to:
\begin{equation}
\omdb = 1 - \frac{\partial \hb }{\partial \hrhodb} \sum_{j} \mj \frac{\partial \kn{b}{j}  \dohb}{\partial \hb}.
\label{eq:omegadb}
\end{equation}
$\tilde{\theta}$ is defined according to
\begin{equation}
\tilde{\theta} \equiv \theta + \frac{\hat{\rho}_{\rm g}}{\hrhod} ( 1 - \theta) (1 - \Omega_{\rm d}) .
\end{equation}
At this stage, no assumptions are made with respect to the functional form of the drag coefficient $K$. The evolution of the internal energy for an SPH gas particle is given by
\begin{align}
\frac{{\rm d}u_{a}}{{\rm d}t}  & = \frac{\tilde{\theta}_{a} P_{a}}{\Omega_{a} \hat{\rho}_{a}^{2} } \sum_{b} m_{b} \left( \va - \vb \right) \cdot \nabla_{a} W_{ab} (h_{a})  \label{eq:en_SPH_gas} \\ \nonumber
& +  \frac{(1 - \theta_{a}) P_{a}}{\hrhoga \hat{\rho}_{{\rm d},a}} \sum^{N_{neigh, dust}}_{j=1} m_{j} \left( {\bf v}_{a} - {\bf v}_{j} \right) \cdot \nabla_{a} W_{aj} (h_{a}) \\ \nonumber
& + \nu \sum_{k}   \mk \frac{\kak}{\hat{\rho}_{a} \hat{\rho}_{k}} \left(\vak \cdot \hrak \right)^{2} D_{ak} (h_{a}).
\end{align}
In Paper~I, we showed that the total linear and angular momentum as well as the total energy are exactly conserved. Thermal coupling terms have been neglected in this paper.

\subsection{Kernel functions}

Two different kernels are employed to perform the SPH interpolations. First, a standard bell-shaped kernel $W$:
\begin{equation}
W\left( r,h\right) = \frac{\sigma}{h^{\nu}} f\left( q \right)  ,
\label{eq:Kerform}
\end{equation}
where $h$ denotes the smoothing lengths of each phases, $\nu$ the number of spatial dimensions and $q\equiv\vert {\bf r} - {\bf r}' \vert / h$ is the dimensionless variable used to calculate the densities and the buoyancy terms. The function $f$ is usually the $M_{4}$ cubic spline kernel \citep{Monaghan2005}. The drag interpolation is performed using a second kernel $D$. As shown in Paper~I,  double-hump shaped kernels given by
\begin{equation}
D\left( r,h\right) = \frac{\tilde{\sigma}}{h^{\nu}} q^{2} f(q),
\label{eq:Kerformdrag}
\end{equation}
significantly improve the accuracy of the drag interpolation --- for the same computational cost --- compared to bell-shaped kernels. The normalisation constants $\tilde{\sigma}$ for various double hump kernels are given in Paper~I. We adopt the double hump cubic for the drag terms in this paper.

\subsection{Astrophysical drag regimes in SPH}

\subsubsection{Gas viscosity and mean free path}

The drag coefficients $K_{ak}$ involved in Eqs.~\ref{eq:mom_SPH_gas} -- \ref{eq:mom_SPH_dust} and \ref{eq:en_SPH_gas} are computed independently for each pair of any gas particle $a$ and dust particle $k$. We first use the sound speed $c_{\rm{s},a}$ to estimate the viscosity $\mu_{a}$ on the gas particle $a$ using (see Eq.~\ref{eq:viscosity_hard})
\begin{equation}
\mu_{a} = \frac{5m}{64 \sigma_{\mathrm{s}}}\sqrt{\frac{\pi}{\gamma}} c_{\rm{s},a} .
\label{eq:viscosity_hard_SPH}
\end{equation}
The mean free path is then computed according to Eq.~\ref{eq:gas_mfp}, giving
\begin{equation}
\lambda_{\mathrm{g},a}  = \dst  \sqrt{\frac{ \pi \gamma }{2 }} \frac{ \mu_{\rm{a}}}{ \hat{\rho}_{a} c_{\rm{s},a}} .
\label{eq:gas_mfp_SPH}
\end{equation}
Finally, $\lambda_{\mathrm{g},a}$ is compared to the quantity $4s_{k}/9$ --- $s_{k}$ being the grain size of the dust particle --- to determine whether the drag coefficient of the SPH pair $K_{ak}$ is calculated using the Epstein or the Stokes drag regimes.

\subsubsection{Epstein regime}

If $4s_{k}/9 \le \lambda_{\mathrm{g},a}$, the drag coefficient $K_{ak}$ is calculated using the Epstein prescription. Introducing the SPH quantity $\psi_{ak}$ calculated on a pair of gas and dust SPH particles and defined by
\begin{equation}
\psi_{ak} \equiv \sqrt{\frac{\gamma}{2}} \frac{\vert {\bf v}_{ak}\vert}{c_{\mathrm{s},a}} ,
\label{eq:def_psi_SPH}
\end{equation}
Eq.~\ref{eq:Epstein_specular} can be straightforwardly translated to get the drag coefficient $K_{ak}$ involved in the SPH drag force
\begin{align}
\kak =  \dst - \sqrt{\pi} s^{2}\rhog \frac{\hrhod}{m_{\rm{d}}} \vert \vak \vert& \left[ \frac{1}{2\sqrt{\pi}} \left\lbrace \left( \frac{1}{\psi_{ak}} + \frac{1}{2 \psi_{ak}^3} \right)e^{-\psi_{ak}^{2}} + \right. \right.  \label{eq:Epstein_specular_SPH} \\  \nonumber
& \left. \left.  \left(  1 + \frac{1}{\psi_{ak}^{2}} - \frac{1}{4 \psi_{ak}^{4}}  \right) \sqrt{\pi} \, \mathrm{erf} \left( \psi_{ak} \right)  \right\rbrace \right]  \textbf{x} ,
\end{align} 
where $s$ is the grain radius, $m_{\rm{d}}$ is the grain mass and $\gamma$ is the adiabatic index. Eq.~\ref{eq:Epstein_specular_SPH} is computationally expensive as it involves exponential and error functions. The SPH equivalent of Eq.\ref{eq:Epstein_mixed} is given by
\begin{equation}
\kak = \frac{4}{3} \pi   \sqrt{\frac{8}{\pi \gamma}} \frac{\hat{\rho}_{k}}{m_{\rm{d}}} \frac{\hrhoga}{\thetaa} s^{2}c_{\mathrm{s},a} \sqrt{1 + \frac{9 \pi}{128} \frac{v_{ak}^{2}}{ c_{\mathrm{s},a}^{2} }  } .
\label{eq:Epstein_mixed_SPH}
\end{equation}
Both Eqs.~\ref{eq:Epstein_specular_SPH} and \ref{eq:Epstein_mixed_SPH} reduce to the linear Epstein regime at low Mach numbers (equivalent of Eq.~\ref{eq:Epstein_linear}) for which the coefficient $\kak$ is
\begin{equation}
\kak = \frac{4}{3} \pi   \sqrt{\frac{8}{\pi \gamma}} \frac{\hat{\rho}_{k}}{m_{\rm{d}}} \frac{\hrhoga}{\thetaa} s^{2}c_{\mathrm{s},a} ,
\label{eq:Epstein_lin_SPH}
\end{equation}
and to the quadratic drag regime at high Mach numbers (equivalent of Eq.~\ref{eq:Epstein_quad}), for which the coefficient $\kak$ is
\begin{equation}
\kak =  \pi   \rhog s^{2} \frac{\hrhod}{m_{\rm{d}}} \vert {\bf v}_{ak} \vert .
\label{eq:Epstein_quad_SPH}
\end{equation}

\subsubsection{Stokes regime}

If $4s_{k}/9 > \lambda_{\mathrm{g},a}$, the drag coefficient $K_{ak}$ is calculated using the Stokes prescription (see Eqs.~\ref{eq:BBO_generalised}--\ref{eq:cd_coefficients}). The local Reynolds number $\Rdak$ is computed for each pair of gas and dust particles using
\begin{equation}
\Rdak \equiv  \frac{2s \hrhoga \left| \vak \right|}{\mu_{a} \thetaa},
\label{eq:def_rdak}
\end{equation}
such that the drag coefficient $K_{ak}$ can be computed according to
\begin{equation}
\kak  =
\begin{cases}
\dst 6 \pi \frac{\hat{\rho}_{k}}{m_{\rm{d}}} \mu_{a} s  & \Rdak <1, \\ 
\dst \frac{12\pi}{2^{0.6}}  \frac{\hat{\rho}_{k}}{m_{\rm{d}}} \frac{\mu_{a}^{0.6}}{\theta_{a}^{0.4} \hat{\rho}_{a}^{0.6}} s^{1.4} \left| \vak \right|^{0.4} & 1<\Rdak< 800, \\
 \dst 0.22 \pi \frac{\hat{\rho}_{k}}{m_{\rm{d}}} \frac{\hat{\rho}_{a}}{\theta_{a}} s^{2} \left| \vak \right| &\Rdak > 800. \\
\end{cases}
\label{eq:exprkak_Sto}
\end{equation}
These expressions have been used by \citet{Ayliffe2011} to compute the drag on planetesimals in a protoplanetary disc.
%
%=======================================================================================================
\section{Timesteping}
\label{sec:timestepping}

\subsection{Explicit timesteping}
\label{sec:explicit}

The simplest method to evolve the evolution equations for the SPH particles is to use an explicit integrator (e.g. the standard Leapfrog). The stability of the system is guaranteed provided the timestep remains smaller than a critical value $\Delta t_{\rm{c}}$. In Paper~I, we  performed a Von Neumann analysis of the continuous equations, deriving the explicit timestepping criterion
\begin{equation}
\Delta t_{\rm{c}, a} = \min_{k} \left[ \frac{\hrhoga \hat{\rho}_{k} }{K_{ak}(\hrhoga + \hat{\rho}_{k} )} \right];  \hspace{0.5cm} \Delta t_{\rm{c}, i} = \min_{b} \left[ \frac{\hrhogb \hat{\rho}_{i} }{K_{bi}(\hrhogb + \hat{\rho}_{i})} \right]; 
\label{eq:crit_laibeSPH}
\end{equation}
for gas and dust particles, respectively, with the minimum being taken over all the particle's neighbours. Although this criterion was derived in Paper~I for linear drag regimes only, it remains valid even for non-linear drag regimes where the drag coefficients depend on the differential velocity between the particles, i.e. $K_{ak} = K_{ak} \left( \vert \mathbf{v}_{ak} \vert \right)$.

\subsection{Implicit timestepping}

When the drag timescale becomes smaller than other time scales in the system (e.g. the Courant condition or the orbital timescale), the timestep restriction of the explicit methods may become prohibitive and implicit methods are required. \citet{Monaghan1997} considered the application of two implicit schemes (the first-order Backward-Euler and second-order Tischer scheme) to SPH dust-gas mixtures. Both schemes are unconditionally stable, but a higher accuracy is achieved with second-order schemes.

\subsubsection{Backward-Euler method}

The Backward-Euler scheme applied to the drag interaction between SPH dust and gas particles is given by
\begin{eqnarray}
\frac{\mathbf{v}^{n+1}_{a} - \mathbf{v}^{n}_{a}}{\Delta t} & = & - \nu \sum_{k} m_{k} \frac{K_{ak}^{n+1}}{\hat{\rho}_{a} \hat{\rho}_{k}} \left(\mathbf{v}^{n+1}_{ak} \cdot \hat{\mathbf{r}}_{ak} \right) \hat{\mathbf{r}}_{ak} D_{ak}  \label{eq:BE_scheme1},\\ 
\frac{\mathbf{v}^{n+1}_{i} - \mathbf{v}^{n}_{i}}{\Delta t} & = & + \nu \sum_{b} m_{b} \frac{K_{bi}^{n+1}}{\hat{\rho}_{b} \hat{\rho}_{i}} \left(\mathbf{v}^{n+1}_{bi} \cdot \hat{\mathbf{r}}_{bi} \right)\hat{\mathbf{r}}_{bi} D_{bi}  \label{eq:BE_scheme2}.
\end{eqnarray}
Although the scheme is unconditionally stable, the implicit equation with respect to the velocities $\textbf{v}^{n+1}$ must be solved at each time step. Direct numerical inversion of this linear system would be prohibitive given the typical number of neighbour interactions for each SPH particle. Thus, approximate or iterative solutions to Eqs. \ref{eq:BE_scheme1} -- \ref{eq:BE_scheme2} are required.

\subsubsection{Monaghan (1997) scheme}
 \citet{Monaghan1997} suggested approximating the velocities $\textbf{v}^{n+1}$ of Eqs. \ref{eq:BE_scheme1} -- \ref{eq:BE_scheme2} using a pairwise treatment in order to preserve the exact conservation of linear and angular momentum in the SPH formalism. Considering the interaction between the SPH gas particle $a$ the dust particle $i$, \citet{Monaghan1997} introduced pairwise auxiliary velocities $\tilde{\bf{v}}$ defined by:
\begin{eqnarray}
\tilde{\textbf{v}}_{a} & = & {\bf v}_{a}^{n}  -   \mi ~ \Delta t \frac{ \nu K_{ai} D_{ai}}{ \hat{\rho}_{a} \hat{\rho}_{i} }  \left( \tilde{\textbf{v}}_{ai} \cdot \hat{\bf{r}}_{ai}  \right ) \hat{\bf{r}}_{ai} ,\label{eq:pairwise_old1}  \\
\tilde{\textbf{v}}_{i}  & = & {\bf v}_{i}^{n}   + \ma ~ \Delta t \frac{ \nu K_{ai} D_{ai}}{ \hat{\rho}_{a} \hat{\rho}_{i} } \left( \tilde{\textbf{v}}_{ai} \cdot \hat{\bf{r}}_{ai}  \right ) \hat{\bf{r}}_{ai} ,
\label{eq:pairwise_old2}
\end{eqnarray}
Eqs.~\ref{eq:pairwise_old1} and \ref{eq:pairwise_old2} are solved, for a given pair of particles, by taking the scalar product by $\hat{\bf{r}}_{ai}$ of the difference of the two equations, giving
\begin{equation}
\tilde{\textbf{v}}_{ai} \cdot \hat{\bf{r}}_{ai} =\frac{ \textbf{v}_{ai}^{n} \cdot \hat{\bf{r}}_{ai} }{1 + \Delta t \frac{ \nu K_{ai} D_{ai}}{ \hat{\rho}_{a} \hat{\rho}_{i} } \left( \ma + \mi \right)} .
\label{eq:vij_starstar_hrij_old}
\end{equation}
Substituting this expression into Eq.~\ref{eq:pairwise_old1} and \ref{eq:pairwise_old2} gives expressions for $\tilde{\textbf{v}}_{a}$ and $\tilde{\textbf{v}}_{i}$. Iterating this pairwise process by looping over all the SPH particles provides an approximate solution for the velocities $\bf{v}^{n+1}$, i.e.
\begin{eqnarray}
\frac{\mathbf{v}^{n+1}_{a} - \mathbf{v}^{n}_{a}}{\Delta t} & \simeq & - \nu \sum_{k} m_{k} \frac{K_{ak}}{\hat{\rho}_{a} \hat{\rho}_{k}} \left(\tilde{\mathbf{v}}_{ak} \cdot \hat{\mathbf{r}}_{ak} \right) \hat{\mathbf{r}}_{ak} D_{ak}  \label{eq:BE_schemeapprox1},\\
\frac{\mathbf{v}^{n+1}_{i} - \mathbf{v}^{n}_{i}}{\Delta t} & \simeq & + \nu \sum_{b} m_{b} \frac{K_{bi}}{\hat{\rho}_{b} \hat{\rho}_{i}} \left(\tilde{\mathbf{v}}_{bi} \cdot \hat{\mathbf{r}}_{bi} \right)\hat{\mathbf{r}}_{bi} D_{bi}  \label{eq:BE_schemeapprox2}.
\end{eqnarray}
The main drawback of this method is that the approximation given by Eqs.~\ref{eq:BE_schemeapprox1} and \ref{eq:BE_schemeapprox2} is inexact -- that is, it provides only an approximate solution to Eqs.~\ref{eq:BE_scheme1} and \ref{eq:BE_scheme2}. Furthermore the accuracy of the approximation is not known \emph{a priori} and there is no possibility of performing repeated sweeps in order to converge to a more accurate solution. In practice, we find that the velocities obtained by this scheme (for example on the \textsc{dustybox} test) can be significantly in error, with no possibility of improving the convergence (for example, by doing several iterations/sweeps).

\subsubsection{Alternative pairwise treatment for linear drag regimes}
\label{sec:newpairwise}
 We propose a more consistent method for solving Eqs.~\ref{eq:BE_scheme1}--\ref{eq:BE_scheme2} on a given gas or dust particle ($a$ and $i$, respectively) by sweeping over all particle pairs and updating the velocities iteratively according to
\begin{eqnarray}
{\textbf{v}}_{a}^{**}  & = & {\bf v}_{a}^{n} + \Delta t \textbf{F}_{a,\mathrm{drag}}^{*} \nonumber \\
&  - &   \mi ~ \Delta t \frac{ \nu K_{ai} D_{ai}}{ \hat{\rho}_{a} \hat{\rho}_{i} }  \left[ \left( {\textbf{v}}_{ai}^{**}  - \textbf{v}_{ai}^{*} \right) \cdot \hat{\bf{r}}_{ai}  \right ] \hat{\bf{r}}_{ai} ,\label{eq:pairwise_new1}  \\
{\textbf{v}}_{i}^{**}   & = & {\bf v}_{i}^{n}  + \Delta t \textbf{F}_{i,\mathrm{drag}}^{*} \nonumber \\
& + & \ma ~ \Delta t \frac{ \nu K_{ai} D_{ai}}{ \hat{\rho}_{a} \hat{\rho}_{i} } \left[ \left( {\textbf{v}}_{ai}^{**} - \textbf{v}_{ai}^{*}\right)\cdot \hat{\bf{r}}_{ai}  \right ] \hat{\bf{r}}_{ai} ,
\label{eq:pairwise_new2}
\end{eqnarray}
where ${\bf v}^{**}$ refers to the improved approximation to ${\bf v}^{n+1}$ obtained after updating each pair and ${\bf v}^{*}$ to the previous iteration value of ${\bf v}^{**}$. Eqs.~\ref{eq:pairwise_new1} and \ref{eq:pairwise_new2} are solved for each pair of particles by taking the dot product of $\hat{\bf{r}}_{ai}$ with the difference of the two equations, giving
\begin{equation}
\left({\textbf{v}}_{ai} \cdot \hat{\bf{r}}_{ai}\right)^{**} =\frac{ \left( \textbf{v}_{ai}^{n} + \Delta t \textbf{F}_{ai,\mathrm{drag}}^{*} + \Delta t \frac{ \nu K_{ai} D_{ai}}{ \hat{\rho}_{a} \hat{\rho}_{i} } \left( \ma + \mi \right)  \textbf{v}_{ai}^{*} \right) \cdot \hat{\bf{r}}_{ai} }{1 + \Delta t \frac{ \nu K_{ai} D_{ai}}{ \hat{\rho}_{a} \hat{\rho}_{i} } \left( \ma + \mi \right)} .
\label{eq:vij_starstar_hrij_new}
\end{equation}
Substituting Eq.~\ref{eq:vij_starstar_hrij_new} in Eqs.~\ref{eq:pairwise_new1} and \ref{eq:pairwise_new2} gives the updated velocities for the pair. Note that during the global sweep over particle pairs ${\bf v}^{*}$ begins as ${\bf v}^{n}$ at the first iteration but is updated as soon as new values become available.

This pairwise correction ensures that i) both the linear and the angular momentum are exactly conserved and ii) the velocities converge to the correct solution of the implicit scheme given by Eqs.~\ref{eq:BE_scheme1} and \ref{eq:BE_scheme2}, since the last term of Eqs.~\ref{eq:pairwise_new1} and \ref{eq:pairwise_new2} tends to zero as the number of iterations increases.
We thus refine our approximation to the solution by performing as many successive iterations as are required to reach a suitable convergence criterion.

\subsubsection{Convergence criterion}

We consider that the approximation we obtain from the implicit scheme described above is accurate enough when
\begin{equation}
\frac{\vert \bf{v}^{k+1} -  \bf{v}^{k} \vert}{\min c_{\rm{s}}} < \varepsilon ,
\label{eq:implicit_crit}
\end{equation}
is satisfied for each particle. Typically, we adopt $\varepsilon = 10^{-4}$, which ensures that the approximation we make on the time stepping is negligible compared to the $\mathcal{O}(h^{2})$ truncation error of the underlying SPH scheme.

\subsubsection{Implementation into Leapfrog}

The Leapfrog scheme is well suited to the evolution of particle methods because, for position-dependant forces, it preserves geometric properties of particle orbits and requires only one evaluation per timestep to give second order accuracy. In the standard formulation, the evolution is computed according to
\begin{equation}
\begin{array}{lrrcll}
{\rm Kick} & \left[ \right. & {\bf v}^{1/2} & = & {\bf v}^{0}  + \frac{\Delta t}{2} {\bf f}^{0} \left({\bf x}^{0} ,{\bf v}^{0}  \right), & \left. \right] \\ [0.5em]
{\rm Drift}  & \left[ \right. & {\bf x}^{1} & = & {\bf x}^{0}  +  \Delta t  {\bf v}^{1/2}, & \left. \right] \\ [0.5em]
{\rm Kick}  & \left[ \right. & {\bf v}^{1} & = & {\bf v}^{1/2} + \frac{\Delta t}{2}   {\bf f}^{1} \left({\bf x}^{1} ,{\bf v}^{1}  \right), & \left. \right] 
\end{array}
\end{equation}
corresponding to Kick, Drift and Kick steps respectively. Adapting Leapfrog to deal with velocity dependent forces (e.g. drag) is \textit{a priori} more difficult since for velocity-dependent forces, the last Kick is implicit in ${\bf v}^{1}$. For our present purposes, this does not present a major problem since the drag is already computed implicitly. Splitting the forces into position-dependent (${\bf f}_{\rm SPH}$) and drag (${\bf f}_{\rm drag}$) contributions, the scheme becomes
\begin{equation}
\begin{array}{lrrcll}
{\rm Kick} & \left[ \right. & {\bf \tilde{v}}^{1/2} & = & {\bf v}^{0}  + \frac{\Delta t}{2} {\bf f}^{0}_{\rm SPH} \left({\bf x}^{0}  \right), & \left. \right] \\ [0.5em]
{\rm Drift}  & \left[ \right. & {\bf x}^{1/2} & = & {\bf x}^{0}  +  \frac{\Delta t}{2}  \tilde{{\bf v}}^{1/2}, & \left. \right] \\ [0.5em]
{\rm Drag} & \left[  \right. & {\bf v}^{1/2} & = & {\bf \tilde{v}}^{1/2}  + \frac{\Delta t}{2} {\bf f}^{1/2}_{\rm drag} \left({\bf x}^{1/2} ,{\bf v}^{1/2}  \right) & \left. \right] \\ [0.5em]
{\rm Drift}  & \left[ \right. & {\bf x}^{1} & = & {\bf x}^{0}  +  \Delta t  {\bf v}^{1/2}, & \left. \right] \\ [0.5em]
{\rm Kick}  & \left[ \right. & {\bf \tilde{v}}^{1} & = & {\bf v}^{1/2}  + \frac{\Delta t}{2} {\bf f}^{1}_{\rm SPH} \left({\bf x}^{1} \right), & \left. \right] \\[0.5em]
{\rm Drag} & \left[  \right. & {\bf v}^{1} & = & {\bf \tilde{v}}^{1}  + \frac{\Delta t}{2} {\bf f}^{1}_{\rm drag} \left({\bf x}^{1} ,{\bf v}^{1}  \right) & \left. \right] 
\end{array}
\label{eq:KDDDKD}
\end{equation}
where the Drag steps represent the implicit updates computed as described in Sec.~\ref{sec:newpairwise}. The disadvantage of Eq.~\ref{eq:KDDDKD} is that two drag force evaluations are required, removing one of the advantages of the Leapfrog integrator. Inspection of \ref{eq:KDDDKD} reveals that an alternative version that requires only one Drag step can be constructed according to
\begin{equation}
\begin{array}{lrlr}
{\rm Kick} &
\bigl[ &
\begin{array}{rcl}
{\bf v}^{1/2} & = & {\bf v}^{0}  + \frac{\Delta t_{0}}{2} {\bf \tilde{f}}, 
\end{array}
 & \bigr]\\ [1.5em]
{\rm Drift} &
\bigl[ &
\begin{array}{rcl}
{\bf x}^{1} & = & {\bf x}^{0}  +  \Delta t_{0}  {\bf v}^{1/2}, 
\end{array}
 & \bigr] \\ [1.em]
{\rm Drag} &
 & \left\lbrace
\begin{array}{rcl}
{\tilde{\bf  v}}^{3/2} & = & {\bf v}^{1/2}  + \frac{\Delta t_{0} + \Delta t_{1} }{2} {\bf f}^{1}_{\rm SPH} \left({\bf x}^{1} \right), \\ [0.5em]
{\bf v}^{3/2} & = &{ \tilde{\bf v}}^{3/2}  + \frac{\Delta t_{0} + \Delta t_{1}}{2} {\bf f}^{1}_{\rm drag} \left({\bf x}^{1} ,{\bf v}^{3/2}  \right), \\ [0.5em]
{\tilde{\bf f}} & = & 2 \left({ \tilde{\bf v}}^{3/2}  - { \tilde{\bf v}}^{1/2} \right) / \left( \Delta t_{0} + \Delta t_{1} \right),
\end{array}
 \right.&\\ [3.em]
{\rm Kick} &
\bigl[ &
\begin{array}{rcl}
{\bf v}^{1} & = & {\bf v}^{1/2}  + \frac{\Delta t_{0}}{2} {\bf \tilde{f}}. 
\end{array}
 & \bigr]
\end{array}
\label{eq:KDDK}
\end{equation}
where we have combined the drag steps by predicting the velocity ${\bf v}^{3/2}$. Note that strictly, the Drag step in this method is semi-implicit since the force is evaluated using ${\bf x}^{1}$ rather than ${\bf x}^{3/2}$. However, we expect this approximation to be reasonable as at high drag (for which the implicit method is designed), the drag mainly changes the differential velocity between the fluids and has less of an effect on the positions. Care is also required when the timestep changes between the steps. We have indicated the correct procedure by specifying $\Delta t_{0}$ and $\Delta t_{1}$ where $\Delta t_{1}$ is the timestep computed based on ${\bf x}^{1}$. Finally, Eq.~\ref{eq:KDDK} requires that ${\tilde{\bf f}}$ is known at the beginning of the integration. This can be easily achieved by performing the Drag step in Eq.~\ref{eq:KDDK} with ${\bf v}^{1/2} = {\bf v}^{0}$, $\Delta t _{0} = 0$ and $\Delta t _{1}$ equal to the timestep calculated using the initial particle positions. 

\subsubsection{Generalisation to non-linear drag regimes}
\label{sec:geneNL}

To extend this alternative pairwise treatment to any non-linear drag regime, two additional points have to be considered. Firstly, although in principle six quantities ($v_{x,y,z}$ for each particle) have  to be determined for each pair,  this can be reduced to a single unknown quantity since the drag coefficient depends only on the modulus of the differential velocity and the exchange of momentum is directed along the line of sight joining the particles. The system of equations for a single pair thus reduces to
\begin{eqnarray}
{\textbf{v}}_{a}^{**}  & = & {\bf v}_{a}^{n} + \Delta t \textbf{F}_{a,\mathrm{drag}}^{*} \nonumber \\
&  - &   \mi ~ \Delta t \frac{ \nu D_{ai}}{ \hat{\rho}_{a} \hat{\rho}_{i} }   K_{ai} \sqrt{ \left[ \left( {\textbf{v}}_{ai}^{**}  - \textbf{v}_{ai}^{*} \right) \cdot \hat{\bf{r}}_{ai} \right]^{2} + V^{2,n}_{\rm orth} } ,\label{eq:pairwiseNL_new1}  \\
{\textbf{v}}_{i}^{**}   & = & {\bf v}_{i}^{n}  + \Delta t \textbf{F}_{i,\mathrm{drag}}^{*} \nonumber \\
& + & \ma ~ \Delta t \frac{ \nu  D_{ai}}{ \hat{\rho}_{a} \hat{\rho}_{i} } K_{ai} \sqrt{ \left[ \left( {\textbf{v}}_{ai}^{**}  - \textbf{v}_{ai}^{*} \right) \cdot \hat{\bf{r}}_{ai} \right]^{2} + V^{2,n}_{\rm orth} } ,
\label{eq:pairwiseNL_new2}
\end{eqnarray}
where
\begin{equation}
V^{2,n}_{\rm orth} =  {\textbf{v}}_{ai}^{n} \cdot  {\textbf{v}}_{ai}^{n}  -  \left( {\textbf{v}}_{ai}^{n} \cdot \hat{\bf{r}}_{ai} \right)^{2} .
\label{eq:defV2}
\end{equation}
Secondly, taking the dot product of $\hat{\bf{r}}_{ai}$ with the difference of the two equations Eqns.~\ref{eq:pairwiseNL_new1} and \ref{eq:pairwiseNL_new2} does not lead in general to an equation which can be solved analytically. The values of ${\textbf{v}}_{ai} \cdot \hat{\bf{r}}_{ai}$ must therefore be determined using a numerical rootfinding procedure (we use a Newton-Raphson scheme) before being substituted in Eqns.~\ref{eq:pairwiseNL_new1} and \ref{eq:pairwiseNL_new2} to determine the velocities for both the gas and the dust particles.

\subsubsection{Performance of the implicit scheme}

The computational cost of a timestep with the implicit pairwise treatment is more expensive than an explicit timestep since at least two iterations have to be performed to ensure that the scheme is converged. However, the implicit pairwise treatment will be more efficient provided that the number of iterations is much smaller than the number of explicit timesteps that would otherwise be required.

We find in practice that the efficiency of the pairwise treatment is mainly determined by the number of iterations required to satisfy Eq.~\ref{eq:implicit_crit} (this aspect was not addressed in the \citet{Monaghan1997} scheme where only one iteration is ever taken in the hope that the approximation is sufficiently accurate). The rapidity of the convergence depends primarily on the ratio $r = \Delta t / t_{\rm s}$ of the timestep over the drag stopping time (defined in Eq.~(96) of Paper~I) and on the value of $\varepsilon$. Empirically, we have found that, for $\varepsilon =10^{-4}$ and $1\lesssim r \lesssim 10$, the implicit pairwise treatment converges efficiently, the ratio $\vert \bf{v}^{k+1} -  \bf{v}^{k} \vert / (\min c_{\rm{s}})$ decreasing by $\sim$ two orders of magnitude at each iterations. Thus, the implicit pairwise treatment improves the computational time by a factor of $\sim 1$--$10$. However, this rapidity of convergence decreases as $r$ increases. At very high drag ($r \gtrsim 1000$), we find that the implicit scheme becomes less efficient than explicit timestepping due to the large number of iterations required. A similar behaviour has been found using the Gauss-Seidel iterative scheme developed by \citet{Whitehouse2005} to treat SPH radiative transfer in the flux-limited diffusion approximation (Bate 2011, private communication), so this issue is not specific to the pairwise treatment.

It is important to note that the computational gain obtained with the pairwise scheme does not solve the resolution issue at high drag extensively discussed in Paper~I. Both of these problems suggest that a more efficient method for handling high drag regimes is required. Such a method is beyond the scope of the present paper.

\subsubsection{Higher order implicit schemes}

Higher temporal accuracy may be achieved by using second instead of first order implicit schemes. The gain in accuracy is obtained by dividing the drag timestep $\Delta t$ into two half timesteps. \citet{Monaghan1997} suggested the `Tischer' scheme, where the two half timesteps are given by
\begin{eqnarray}
\frac{\mathbf{v}^{n+\frac{1}{2}}_{a} - \mathbf{v}^{n}_{a}}{\Delta t/2} & = & - 0.6 \nu \sum_{k} m_{k} \frac{K_{ak}}{\hat{\rho}_{a} \hat{\rho}_{k}} \left(\mathbf{v}^{n+\frac{1}{2}}_{ak} \cdot \hat{\mathbf{r}}_{ak} \right) \hat{\mathbf{r}}_{ak} D_{ak}  ,\nonumber \\
&& -0.4 \nu \sum_{k} m_{k} \frac{K_{ak}}{\hat{\rho}_{a} \hat{\rho}_{k}} \left(\mathbf{v}^{n}_{ak} \cdot \hat{\mathbf{r}}_{ak} \right) \hat{\mathbf{r}}_{ak} D_{ak} , \label{eq:Tischer1_scheme1} \\
\frac{\mathbf{v}^{n+\frac{1}{2}}_{i} - \mathbf{v}^{n}_{i}}{\Delta t/2} & = & +0.6 \nu \sum_{b} m_{b} \frac{K_{bi}}{\hat{\rho}_{b} \hat{\rho}_{i}} \left(\mathbf{v}^{n+\frac{1}{2}}_{bi} \cdot \hat{\mathbf{r}}_{bi} \right)\hat{\mathbf{r}}_{bi} D_{bi}  , \nonumber \\
&& + 0.4 \nu \sum_{b} m_{b} \frac{K_{bi}}{\hat{\rho}_{b} \hat{\rho}_{i}} \left(\mathbf{v}^{n}_{bi} \cdot \hat{\mathbf{r}}_{bi} \right)\hat{\mathbf{r}}_{bi} D_{bi} , \label{eq:Tischer1_scheme2} 
\end{eqnarray}
and then
\begin{eqnarray}
\frac{\mathbf{v}^{n+1}_{a} - \left[ 1.4\mathbf{v}^{n+\frac{1}{2}}_{a}  - 0.4\mathbf{v}^{n}_{a}\right]}{\Delta t/2}& = & - 0.6 \nu \sum_{k} m_{k} \frac{K_{ak}}{\hat{\rho}_{a} \hat{\rho}_{k}} \left(\mathbf{v}^{n+1}_{ak} \cdot \hat{\mathbf{r}}_{ak} \right) \hat{\mathbf{r}}_{ak} D_{ak} , \nonumber \\ \label{eq:Tischer2_scheme1} \\ 
\frac{\mathbf{v}^{n+1}_{i} - \left[1.4 \mathbf{v}^{n+\frac{1}{2}}_{i}  - 0.4\mathbf{v}^{n}_{i} \right]}{\Delta t/2} & = & +0.6 \nu \sum_{b} m_{b} \frac{K_{bi}}{\hat{\rho}_{b} \hat{\rho}_{i}} \left(\mathbf{v}^{n+1}_{bi} \cdot \hat{\mathbf{r}}_{bi} \right)\hat{\mathbf{r}}_{bi} D_{bi}. \nonumber \\ \label{eq:Tischer2_scheme2}
\end{eqnarray}
The last terms of Eqs.~\ref{eq:Tischer1_scheme1}--\ref{eq:Tischer1_scheme2} correspond to the explicit drag force involved in the Forward-Euler scheme. These quantities are computed form the velocities at the timestep $n$ at the beginning of the scheme, as described in Sec.~\ref{sec:explicit}. Then, the successive determination of $\bf{v}^{n+\frac{1}{2}}$ and $\bf{v}^{n}$ as given by Eqs.~\ref{eq:Tischer2_scheme1}--\ref{eq:Tischer2_scheme2}  consists of two Backward-Euler steps with step size $\frac{\Delta t}{2}$. They are therefore computed using our alternative pairwise scheme, until the iterations for each half time step have converged. Eqs.~\ref{eq:Tischer1_scheme1} and \ref{eq:Tischer1_scheme2} concerns the specific case of a linear drag regime, but this scheme can easily be extended to non-linear drag regimes as in Sec.~\ref{sec:geneNL}.

%=======================================================================================================
\section{Numerical tests}
\label{sec:tests}

\subsection{\textsc{dustybox}: Two fluid drag in a periodic box}

The \textsc{dustybox} problem presented by \citet{LP11a} and described in detail in Paper~I consists of two fluids in a periodic box moving with a differential velocity ($\Delta v _{0} =  v _{d,0} -  v _{g,0}$). This is the only test where analytic solutions are known for several functional forms corresponding to non-linear drag regimes (see \citealt{LP11a}). These represent the functional forms of the Epstein and Stokes prescription. We thus use the \textsc{dustybox} problem to benchmark the accuracy of our algorithm for non-linear drag regimes using both explicit and implicit timestepping.

\subsubsection{\textsc{dustybox}: setup}

We set up constant densities $\hrhog$ and $\hrhod$ and gas pressure $\Pg$ in a 3D periodic domain $x, y, z \in [0,1]$ filled by $20^{3}$ gas particles set up on a regular cubic lattice and $20^{3}$ dust particles shifted by half of the lattice spacing in each direction (as in Paper~I, we verified that the results are independent of the offset of the dust lattice). The gas sound speed, the gas and the dust densities are set to unity in code units and no artificial viscosity is applied. The intrinsic dust volume is neglected by assuming $\theta = 1$.

Simulations have been performed using both the explicit timestepping presented in Paper~I and the implicit pairwise timestepping described in Sec.~\ref{sec:timestepping}. For the latter, we verified that both the total linear and angular momentum are exactly conserved as expected.

\subsubsection{\textsc{dustybox}: different drag regimes}
\begin{figure}
   \centering
   \includegraphics[angle=0, width=\columnwidth]{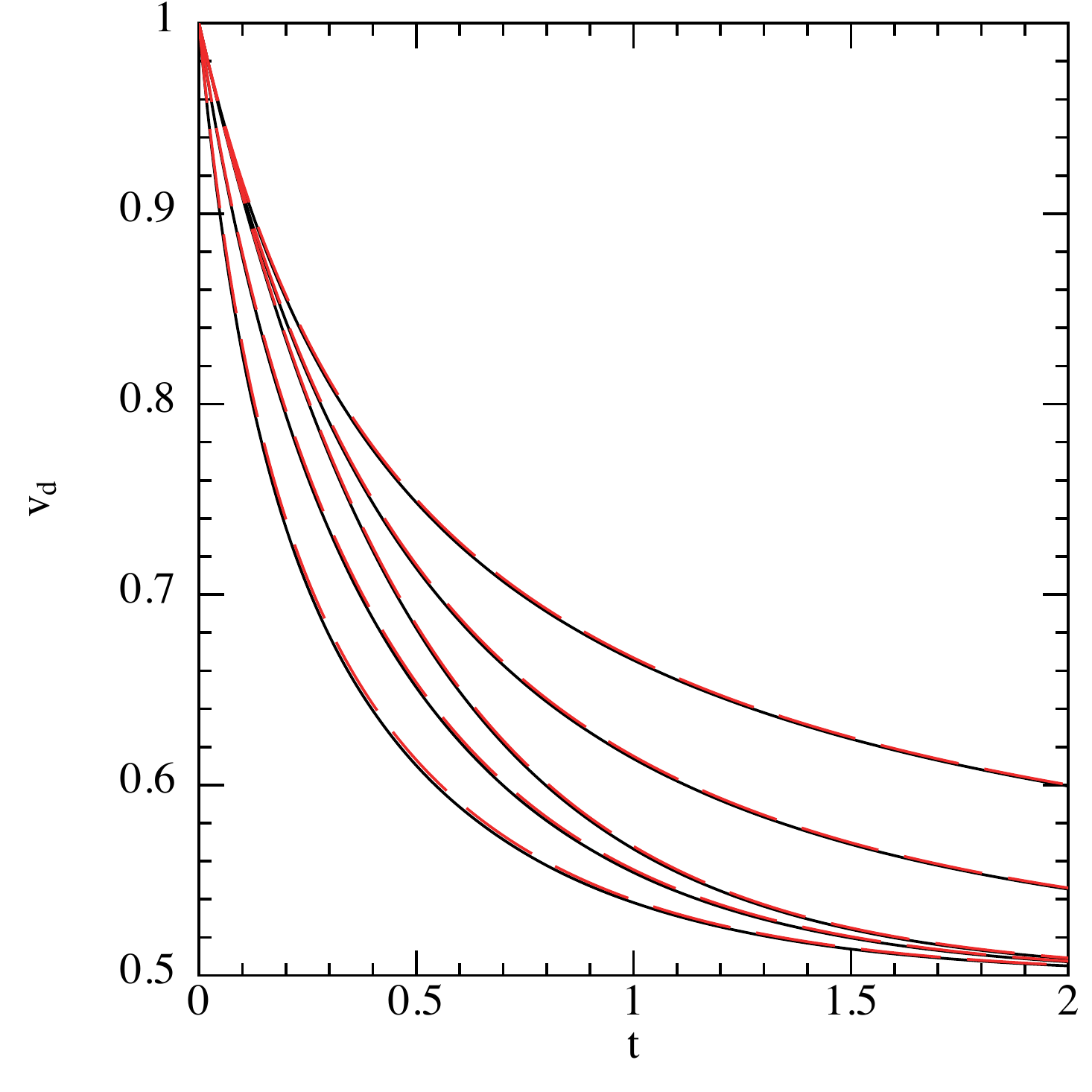} 
   \caption{Dust velocity (solid lines) as a function of time in the \textsc{dustybox} test, using $2 \times 20^{3}$ particles, a dust-to-gas ratio of unity and five different linear and non-linear drag regimes --- quadratic, power-law, linear, third order expansion and mixed, from top to bottom --- compared to the exact solution for each case (long dashed/red lines). The initial velocities are set to $\vdi = 1$, $\vgi = 0$ and the time integration is performed using using the pairwise implicit treatment described in Sec.~\ref{sec:timestepping}. The accuracy ($\lesssim 0.1\%$) of the SPH treatment for dust-gas mixtures is obtained by using the double-hump cubic kernel.}
   \label{fig:NLbox}
\end{figure}
Fig.~\ref{fig:NLbox} shows the results of the \textsc{dustybox} test for the five different regimes given in Table~1 of \citet{LP11a}: linear ($K = K_{0}$), quadratic ($K = K_{0} \vert \Delta v \vert$), power-law ($K = K_{0} \vert \Delta v \vert ^{a}$, with $a = 0.4$), third order expansion ($K = K_{0} [1 + a_{3} \vert \Delta v \vert^{2}]$, with $a_{3} = 0.5$) and mixed ($K = K_{0} \sqrt{1 + a_{2} \vert \Delta v \vert^{2}}$, with $a_{2} =5$) where we have used $K_{0} = 1$ in each case. The analytic solutions are reproduced within an accuracy comprised between $0.1\%$ and $1\%$ in every case --- both linear and non-linear. The implicit scheme was find to converge quickly for this problem, requiring no more than two iterations at every stage of the evolution in each case.

The efficiency of the damping for the \textsc{dustybox} problem decreases when the exponent of the drag regime increases, since $\vert| \Delta v \vert| < 1$. On the contrary, additional non-linear terms give an additional contribution to the drag for the mixed and third order drag regimes, leading to a differential velocity that is more efficiently damped compared to the linear case.

\subsection{\textsc{dustywave}: Sound waves in a dust-gas mixture}

The exact solution for linear waves propagating in a dust-gas mixture (\textsc{dustywave}) was derived in \citet{LP11a} assuming a linear drag regime. Unfortunately, exact solutions prove difficult to obtain for the case of non-linear drag. Instead, we have verified simply that our simulations of the \textsc{dustywave} problem for non-linear drag regimes are converged in both space and time with an explicit timestepping scheme. We have then used these results to benchmark our simulations using implicit timestepping.

We run the \textsc{dustyywave} problem for the same five non-linear drag regimes used above. Strictly speaking, It should be noted that assuming $K_{0}$ is constant (which we assume for this test problem) corresponds to Epstein and Stokes drag only to first order for the \textsc{dustywave} problem. 

\subsubsection{\textsc{dustywave}: Setup}

The \textsc{dustywave} test is performed in a 1D periodic box, placing equally spaced particles in the periodic domain $x \in [0,1]$ such that the gas and dust densities are unity in code units. We do not apply any form of viscosity and the gas sound speed is set to unity. To remain in the linear acoustic regime, the relative amplitude of the perturbation of both velocity and density are set to $10^{-4}$.

\subsubsection{\textsc{dustywave}: Different drag regimes}
\begin{figure}
   \centering
   \includegraphics[angle=0, width=\columnwidth]{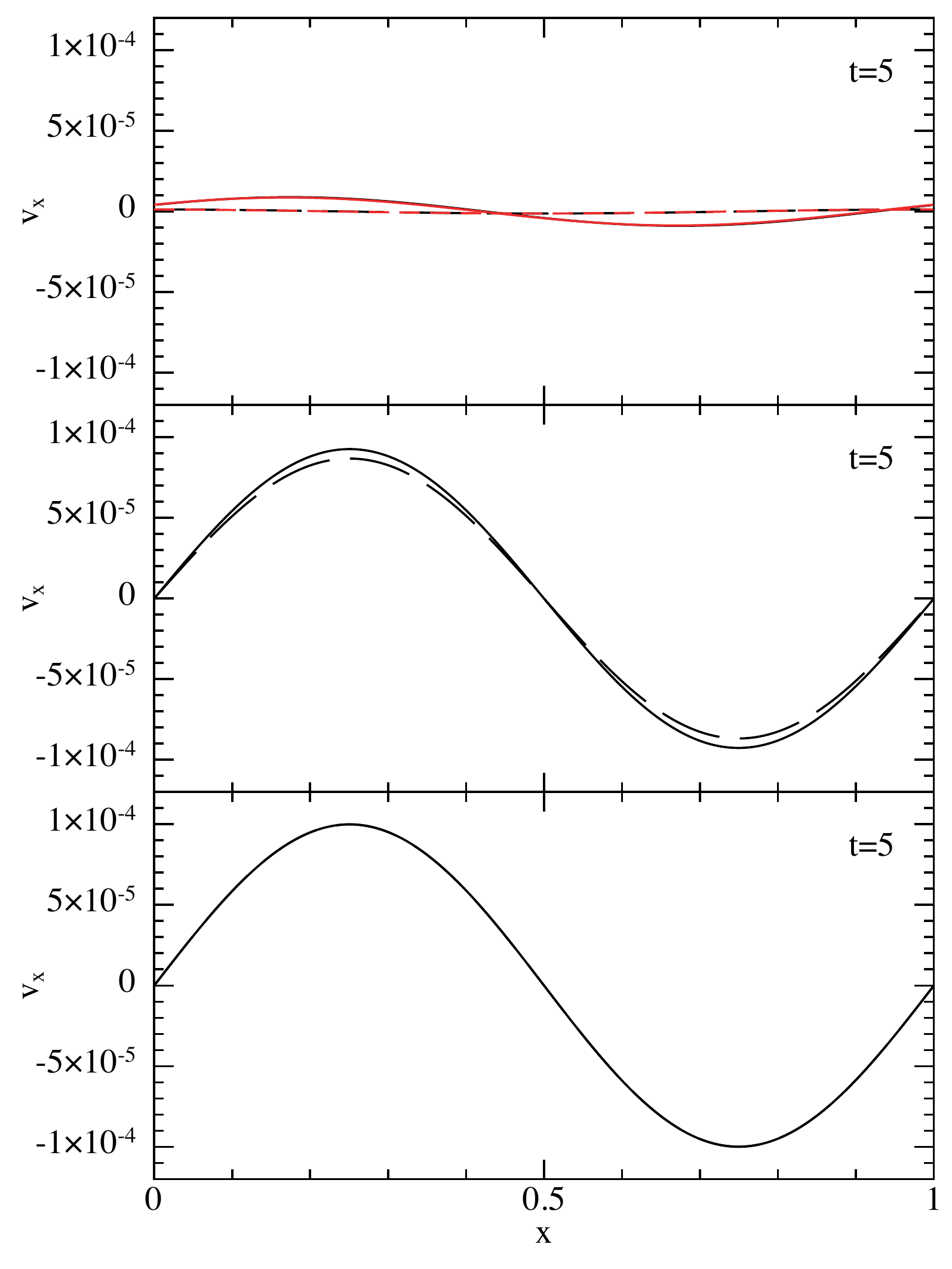} 
   \caption{Solution of the 1D \textsc{dustywave} problem showing the SPH gas (solid black lines) and dust (dashed black) velocities using three different drag regimes: linear (top panel), power-law with an exponent of $0.4$ (center panel) and quadratic (bottom panel). The damping is strongly reduced for non-linear drag regimes compared to the linear case. Results are shown after 5 wave periods and the linear case (top panel) may be compared to the exact analytic solution (red lines). The dust to gas ratio and $K_{0}$ are set to 1 in code units.}
   \label{fig:waveNL}
\end{figure}

Fig.~\ref{fig:waveNL} shows the velocity profiles after 5 periods for three drag regimes --- linear, power law and quadratic --- in the 1D \textsc{dustywave} problem using $K_{0} = 1$ and a dust-to-gas ratio of unity. The solution obtained for the linear drag regime shows an efficiently damped perturbation at $t = 5$, consistent with a stopping time of order unity. By comparison, the perturbation is only weakly damped for the case of the power-law drag regime at the same time. The drag is weaker still in the quadratic regime, for which both the dust and the gas are mostly decoupled. Indeed, since the drag stopping time is a decreasing function of the differential velocity for non-linear drag, and the differential velocity is small with respect to the sound speed, the damping is inefficient. 

We have also performed \textsc{dustywave} simulations using the third order expansion and mixed drag regimes. However, for these cases, non-linear terms represent only negligible corrections compared to the linear term, thus giving the same results as for the linear case.

\subsection{\textsc{dustyshock}: Two fluid dust-gas shocks}

The \textsc{dustyshock} problem (Paper~I) is a two fluid version of the standard \citet{sod78} shock tube problem. Results were presented in Paper~I using a constant drag coefficient $K$ and no heat transfer between the gas and the dust phase. Here we extend the test to non-linear drag regimes. While the evolution during the transient stage is dependent on the drag regime, the solution during the stationary stage remains unchanged, being a fixed function of the gas sound speed and the dust-to-gas ratio. To facilitate the comparison between a physical Epstein drag and the case of a constant drag coefficient (Paper~I), we fix the ratio $s^{2}/m_{\rm{d}}$ --- $s$ being the grain size and $m_{\rm{d}}$ the grain mass--- to unity in code units, such that the Epstein drag coefficient is unity in regions where $\hrhog = 1$.

\subsubsection{\textsc{dustyshock}: Setup}

Equal mass particles are placed in the 1D domain $x \in [-0.5, 0.5]$, where for $x<0$ we use $\rhog = \rhod = 1$, $v_{\rm{g}} = v_{\rm{d}} = 0$ and $P_{\rm{g}} = 1$, while for $x>0$ $\rhog = \rhod = 0.125$, $v_{\rm{g}} = v_{\rm{d}} = 0$ and $P_{\rm{g}} = 0.1$. We use an ideal gas equation of state $P = (\gamma - 1)\rho u$ with $\gamma = 5/3$. Initial particle spacing to the left of the shock in both fluids is $\Delta x = 0.001$ while to the right it is $\Delta x = 0.008$, giving $569$ equal mass particles in each phase. Standard SPH artificial viscosity and conductivity terms are applied as in Paper~I.

\subsubsection{\textsc{dustyshock}: Different drag regimes}

\begin{figure*}
   \centering
   \includegraphics[angle=0, width=\columnwidth]{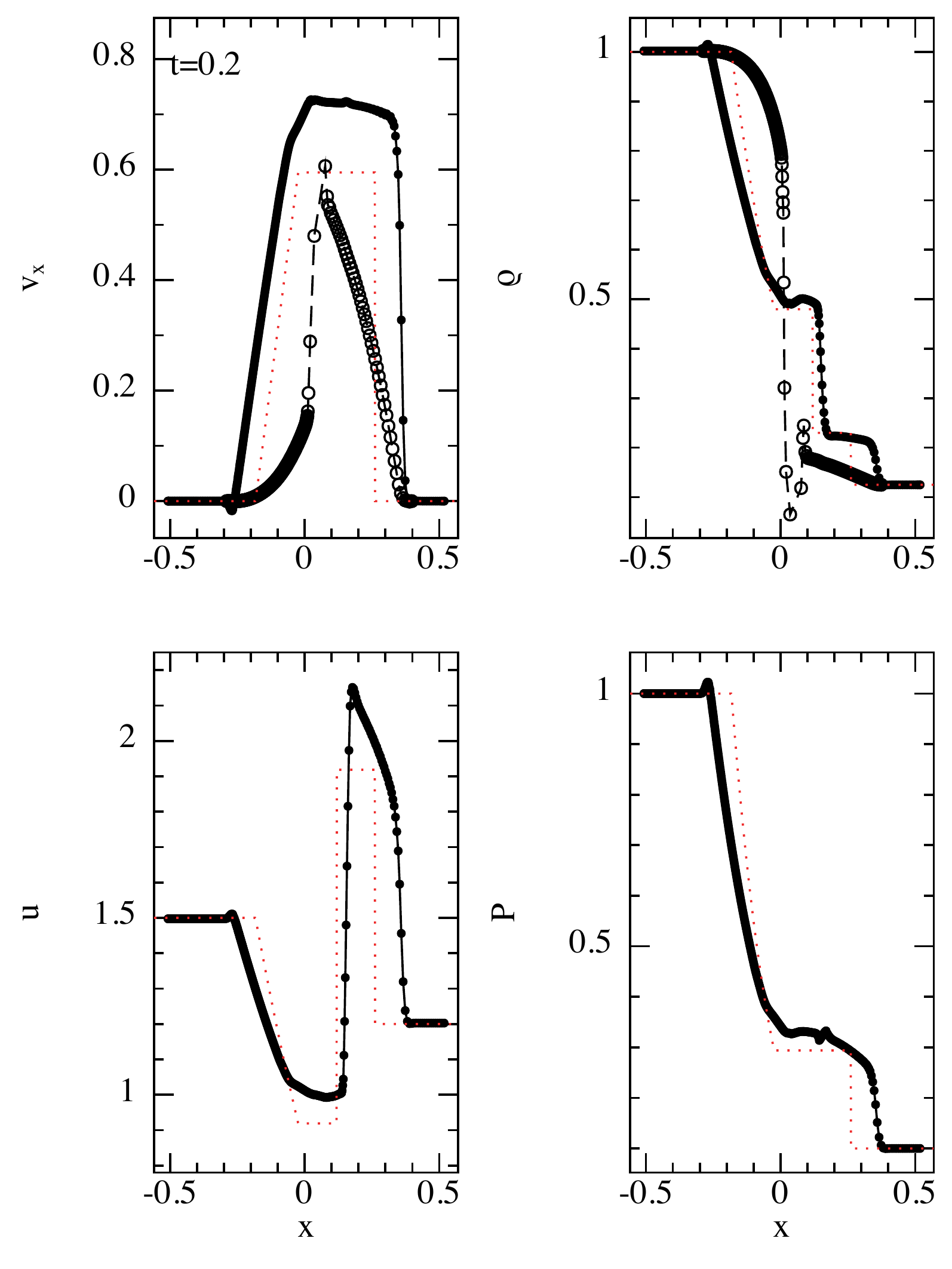} 
   \hspace{0.5cm}
   \includegraphics[angle=0, width=\columnwidth]{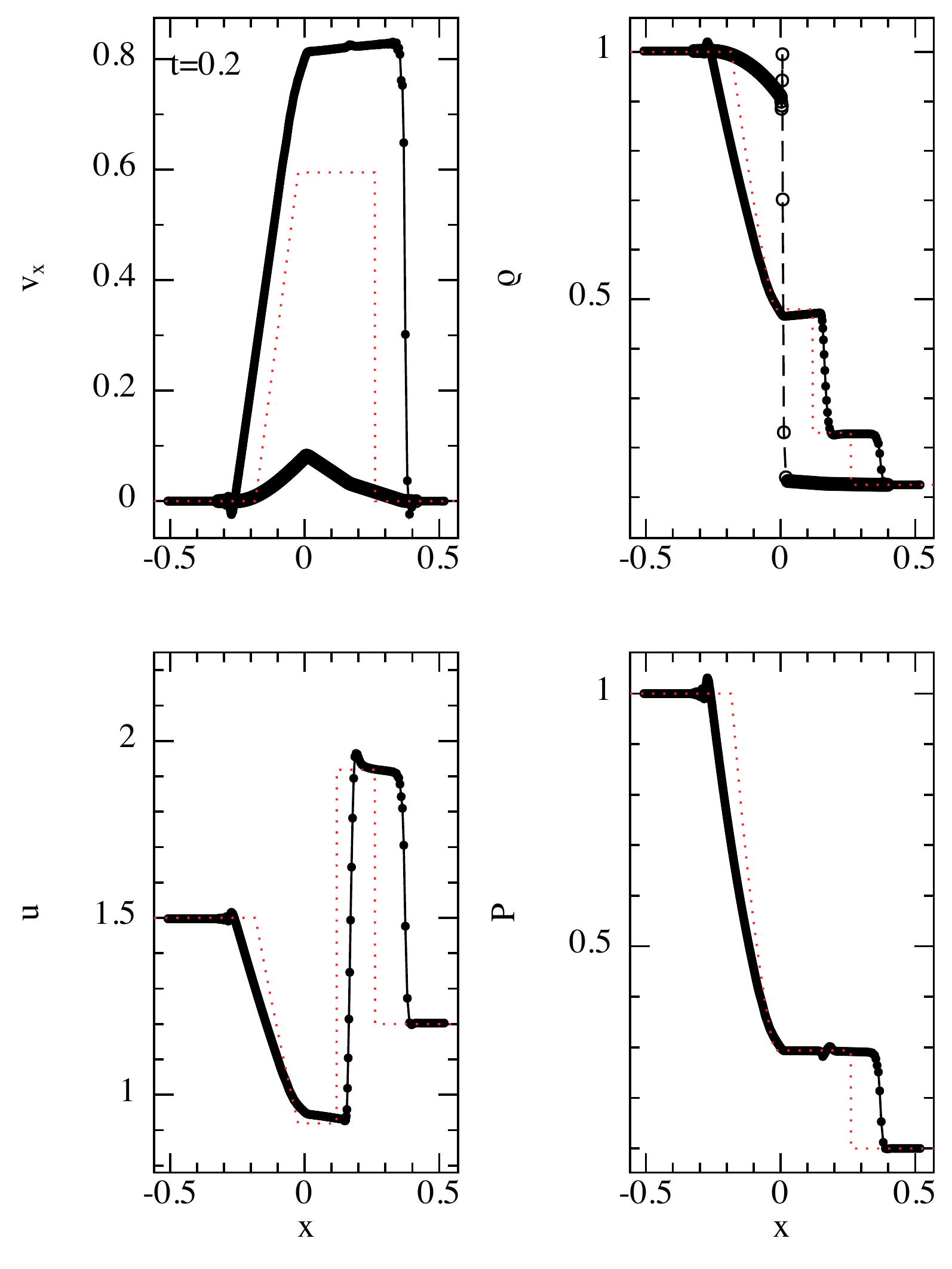} 
   \caption{Results of the \textsc{dustyshock} problem with a linear constant drag coefficient ($K=1$) (left panel) and a non-linear Epstein drag (right panel) where the drag coefficients are initially the same ahead of the shock. The dust-to-gas ratio is set to unity. At $t=0.2$, the solutions are in the transient stage where the analytic solution is not known. As an indication, the solution for the later stationary stage is shown by the dotted/red lines. The profiles differ as the damping is less efficient using the non-linear Epstein regime.}
   \label{fig:shock_NL_Epst}
\end{figure*}

Fig.~\ref{fig:shock_NL_Epst} illustrates how the transient regime of the \textsc{dustyshock} is affected when treating the drag with an astrophysical prescription where the drag coefficient depends on the local density and the gas sound speed (right panel) rather than a constant coefficient (left panel). Specifically, we use the non-linear Epstein drag regime given by Eq.~\ref{eq:Epstein_mixed_SPH}. In this \textsc{dustyshock} test, the non-linear terms constitute a small correction to the linear Epstein drag regime given by Eq.~\ref{eq:Epstein_lin_SPH}. Fig.~\ref{fig:shock_NL_Epst} shows that in the case of the Epstein regime, the drag is less efficient than for the constant coefficient case, leading to a larger ($\sim$ by a factor $7$) differential velocity between the gas and the dust after $t=0.2$ in code units. The dust velocity profile is also smoother than in the constant coefficient case. As a result, the kinetic energy is less efficiently dissipated by the drag, leading to a less sharp peak in the internal energy of the gas. The density profile of the dust is also closer to its initial profile behind and ahead of the shock.

\subsection{\textsc{dustysedov}: Two fluid dust-gas blast wave}

The \textsc{dustysedov} problem (Paper~I) involves the propagation of a blast wave in an astrophysical mixture of dust and gas. We adopt physical units for this problem, assuming a box size of $1$ pc, an ambient sound speed of $2 \times 10^{4}$ cm/s and a gas density of $\rho_{0} = 6 \times 10^{-23}$ g/cm$^{3}$ the energy of the blast is $2 \times 10^{51}$ erg and time is measured in units of $100$ years, roughly corresponding to a supernova blast wave propagating into the interstellar medium. We these units, we choose the grain size, $0.1 \mu$m and the dust-to-gas ratio, $0.01$, to be typical of the interstellar medium. In code units, this corresponds to an initial drag coefficient of $K = 1$ outside the blast radius. As for the \textsc{dustyshock}, we compare the results using a non-linear Epstein drag prescription with with the constant coefficient case described in Paper~I.

\subsubsection{\textsc{dustysedov}: Setup}

The problem is set up in a 3D periodic box ($x,y,z \in [-0.5, 0.5]$), filled by $50^{3}$ particles for both the gas and the dust. Gas particles are set up on a regular cubic lattice, with the dust particles also on a cubic lattice but shifted by half of the lattice step in each direction. For shock-capturing, we set $\alpha_{\rm SPH} = 1$ and  $\beta_{\rm SPH} = 2$ for the artificial viscosity terms and $\alpha_{u} = 1$ for the artificial conductivity term. An ideal gas equation of state $P = (\gamma -1)\rho u$ is adopted with $\gamma = 5/3$.

The internal energy is distributed of the gas over the particles located inside a radius $r < r_{\rm{b}}$ where $r_{\rm b}$ is set to 2h (i.e., the radius of the smoothing kernel which for $50^{3}$ particles and $\eta = 1.2$ is $0.048$). In code units the total blast energy is $E=1$, with $\hrhog = 1$ and $\hrhod = 0.01$. For $r>r_{\rm{b}}$, the gas sound speed is set to be $2 \times 10^{-5}$ in code units.

\subsubsection{\textsc{dustysedov}: Different drag regimes}

\begin{figure}
   \centering
   \includegraphics[angle=0, width=0.49\columnwidth]{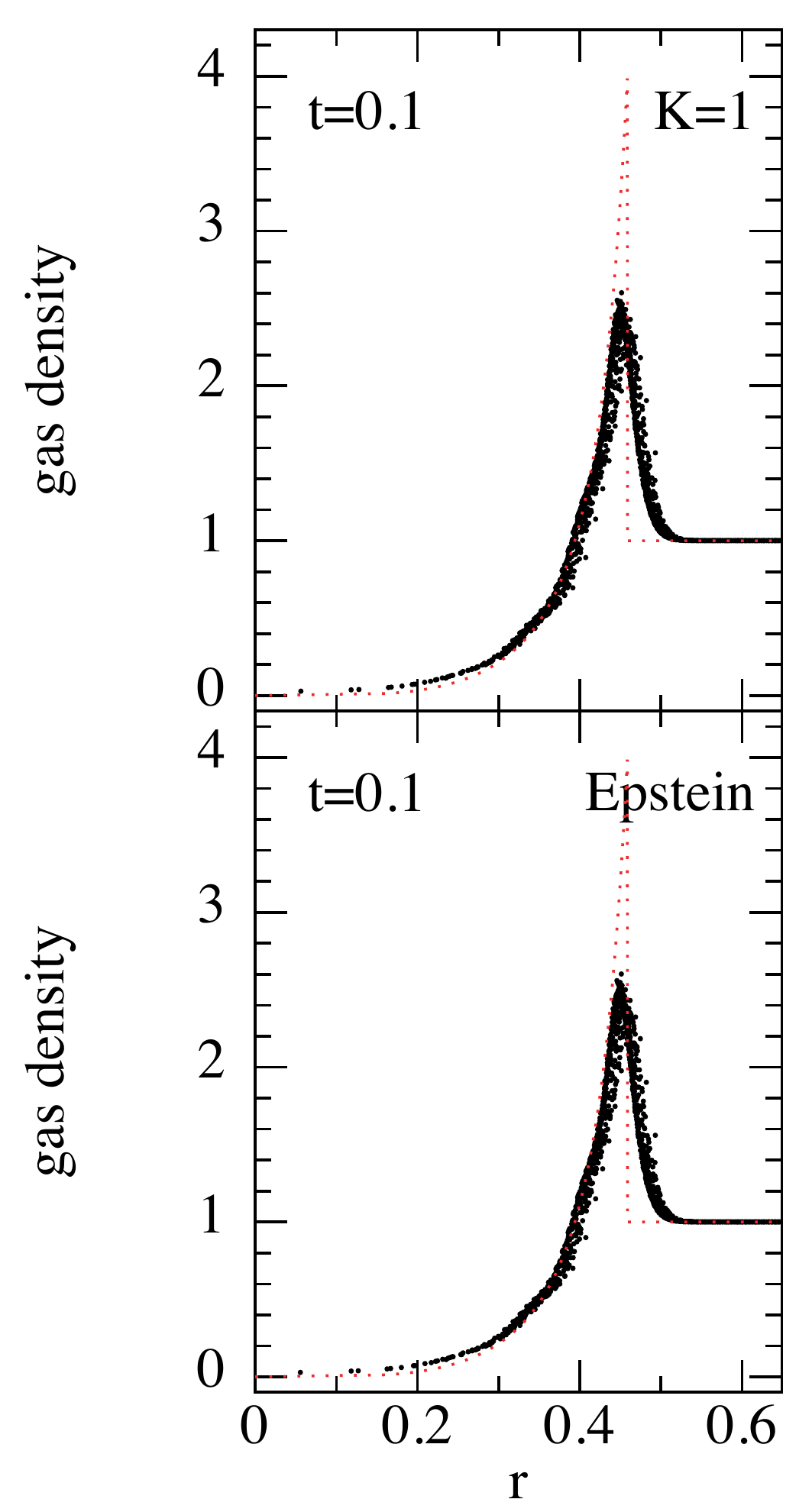} 
   \includegraphics[angle=0, width=0.49\columnwidth]{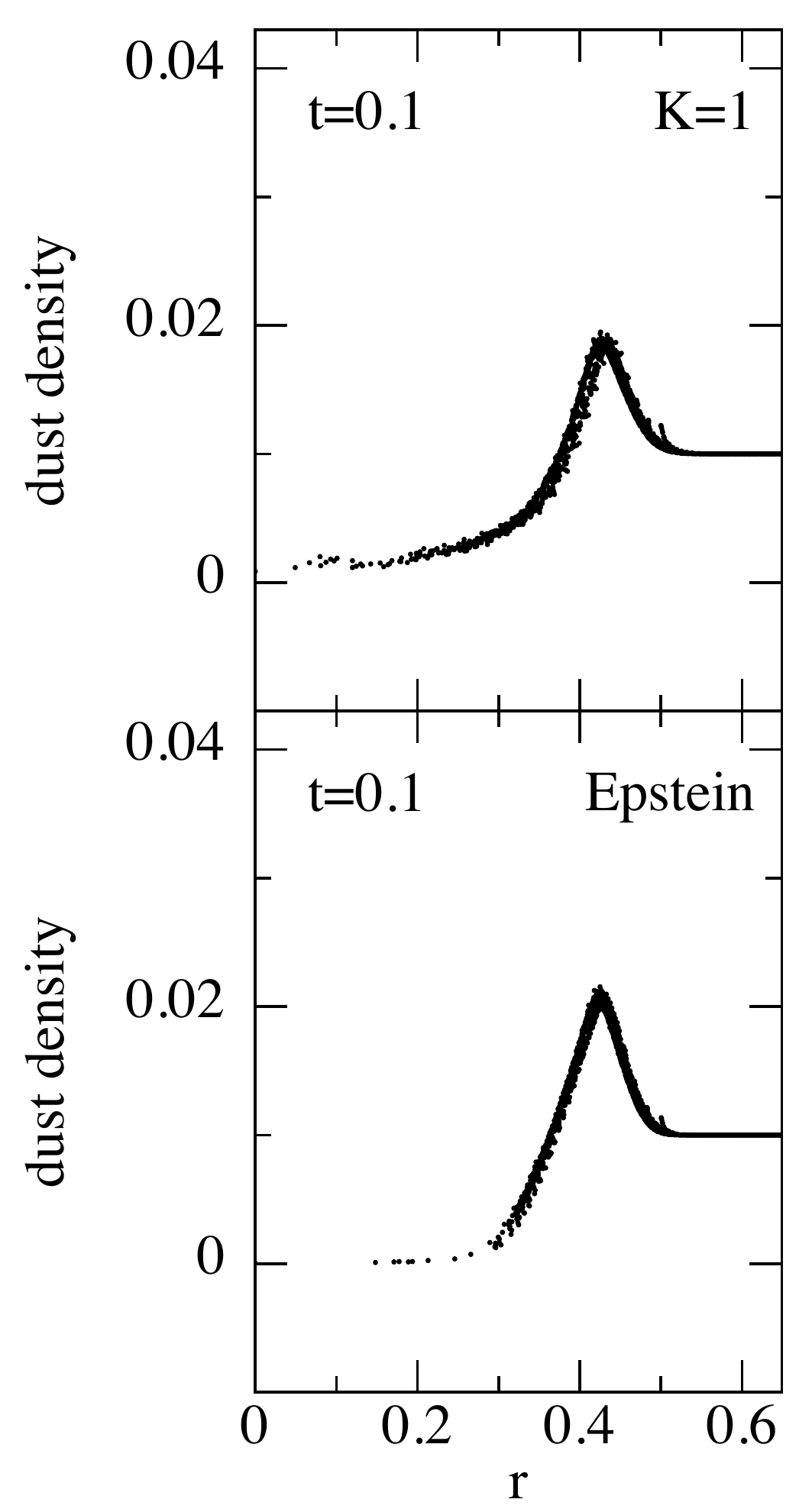} 
   \caption{Results of the 3D \textsc{dustysedov} test, showing the density in the gas (left figure) and dust (right figure) from a Sedov blast wave propagating in an astrophysical ($1\%$ dust-to-gas ratio) mixture of gas and $0.1 \mu$m dust grains in a $1$ pc box.  The drag coefficient is constant ($K = 1$, top panels) or given by the Epstein regime (bottom panels). The low dust-to-gas ratio means that the gas is only weakly affected by the drag from the dust, and is thus close to the self-similar Sedov solution (dotted/red line). In the Epstein case, the drag is much higher inside the blast radius and the dust particles are efficiently piled up by the passage of the gas over-density.}
   \label{fig:sedov_epstein_density}
\end{figure}

\begin{figure*}
   \centering
   \includegraphics[angle=0, width=0.9\textwidth]{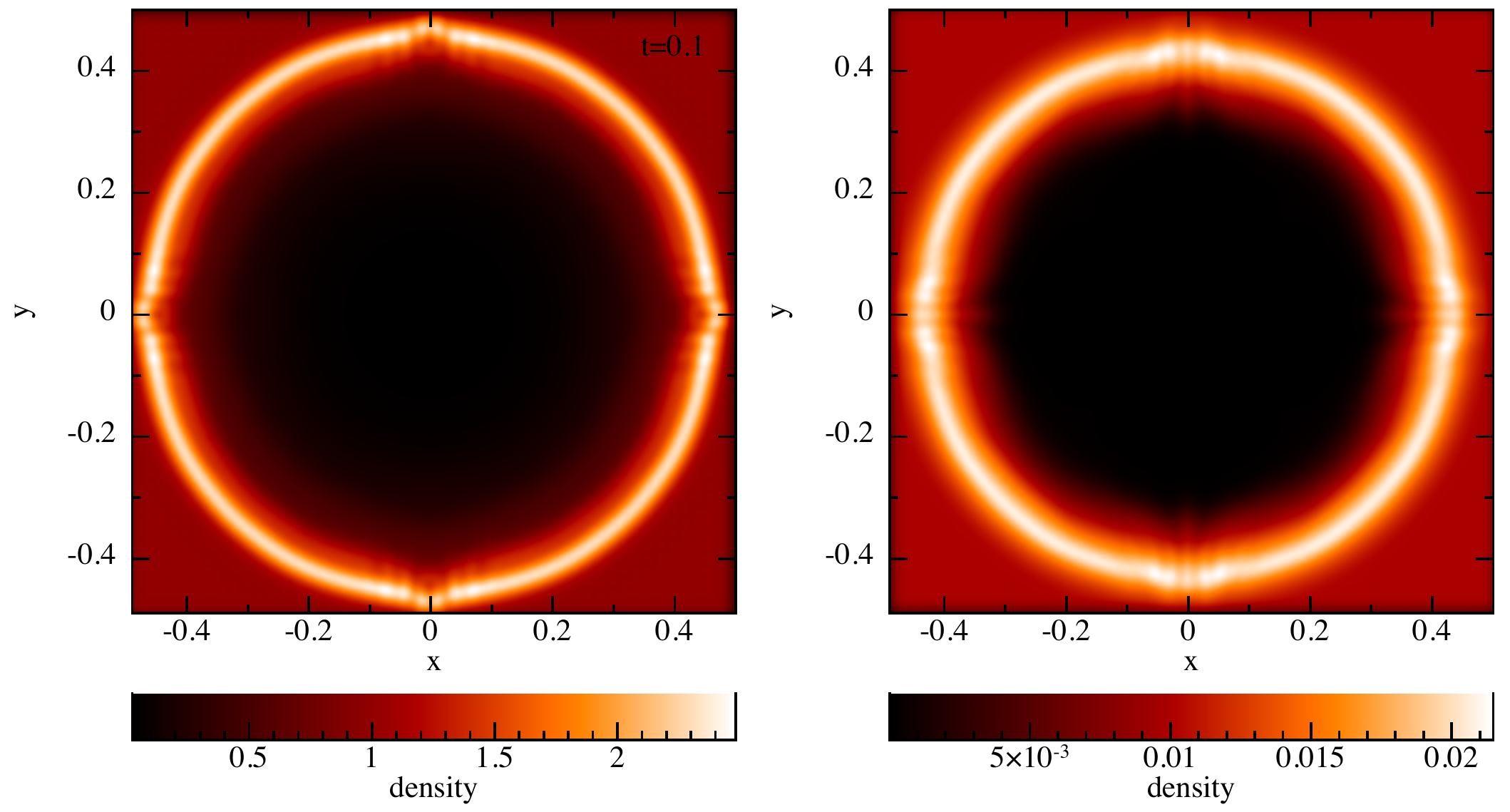} 
   \caption{Cross-section slice showing density in the midplane in the 3D \textsc{dustysedov} problem, for both the gas (left panel) and the dust (right panel) at $t=0.1$. Initially, the dust-to-gas ratio is $0.01$ and the drag coefficient is given by the Epstein regime for grains of $0.1 \mu$m in size. $50^{3}$ SPH particles have been used in each phase.}
   \label{fig:render_epstein}
\end{figure*}

Figs.~\ref{fig:sedov_epstein_density} and \ref{fig:render_epstein} show the evolution of the gas and dust mixture where a constant drag coefficient is used (top panels of Fig.~\ref{fig:sedov_epstein_density}) compared to a drag prescribed by the Epstein regime (bottom panels of Fig.~\ref{fig:sedov_epstein_density}, Fig.~\ref{fig:render_epstein}). The gas profiles are similar in both cases since the gas is poorly affected by the dust given the low dust-to-gas ratio. However, the dust density profiles differ, essentially due to the fact that the drag coefficient scales with the sound speed and is thus higher in the inner blast region for the Epstein case. Thus, the dust is efficiently piled up and accumulates in the gas over-density. As a result, the dust is cleaned up by the gas in the inner regions of the blast, but is more concentrated (by $\sim 10\%$) close to the gas over-density than for the constant drag coefficient case.

The results using either explicit or implicit timestepping were found to be indistinguishable. For the Epstein case, we found that roughly ten iterations were required for the implicit scheme to converge on this problem.

\subsection{\textsc{dustydisc}}

The \textsc{dustydisc} problem concerns the evolution of a dusty gas mixture in a protoplanetary disc (see Paper~I for details). For our test case, we study how the dust distribution is affected when considering a general non-linear Epstein drag instead of the standard linear regime. The results obtained when implicitly integrating the non-linear drag regime have been found to be similar to benchmark tests performed with explicit integration.

\subsubsection{\textsc{dustydisc}: Setup}
We setup $10^{5}$ gas particles and $10^{5}$ dust particles in a $0.01 M_{\odot}$ gas disc (with $0.0001 M_{\odot}$ of dust) surrounding a $1 M_{\odot}$ star. The disc extends from 10 to 400 AU. Both gas and dust particles are placed using a Monte-Carlo setup such that the surface density profiles of both phases are $\Sigma \left( r \right) \propto r^{-1}$. The radial profile of the gas temperature is taken to be $T\left( r \right) \propto r^{-0.6}$ with a flaring $H/r = 0.05$ at 100 AU. One code unit of time corresponds to $10^{3}$ yrs.

\subsubsection{\textsc{dustydisc}: Evolution of the particles}

\begin{figure*}
   \centering
   \includegraphics[angle=0, width=0.9\textwidth]{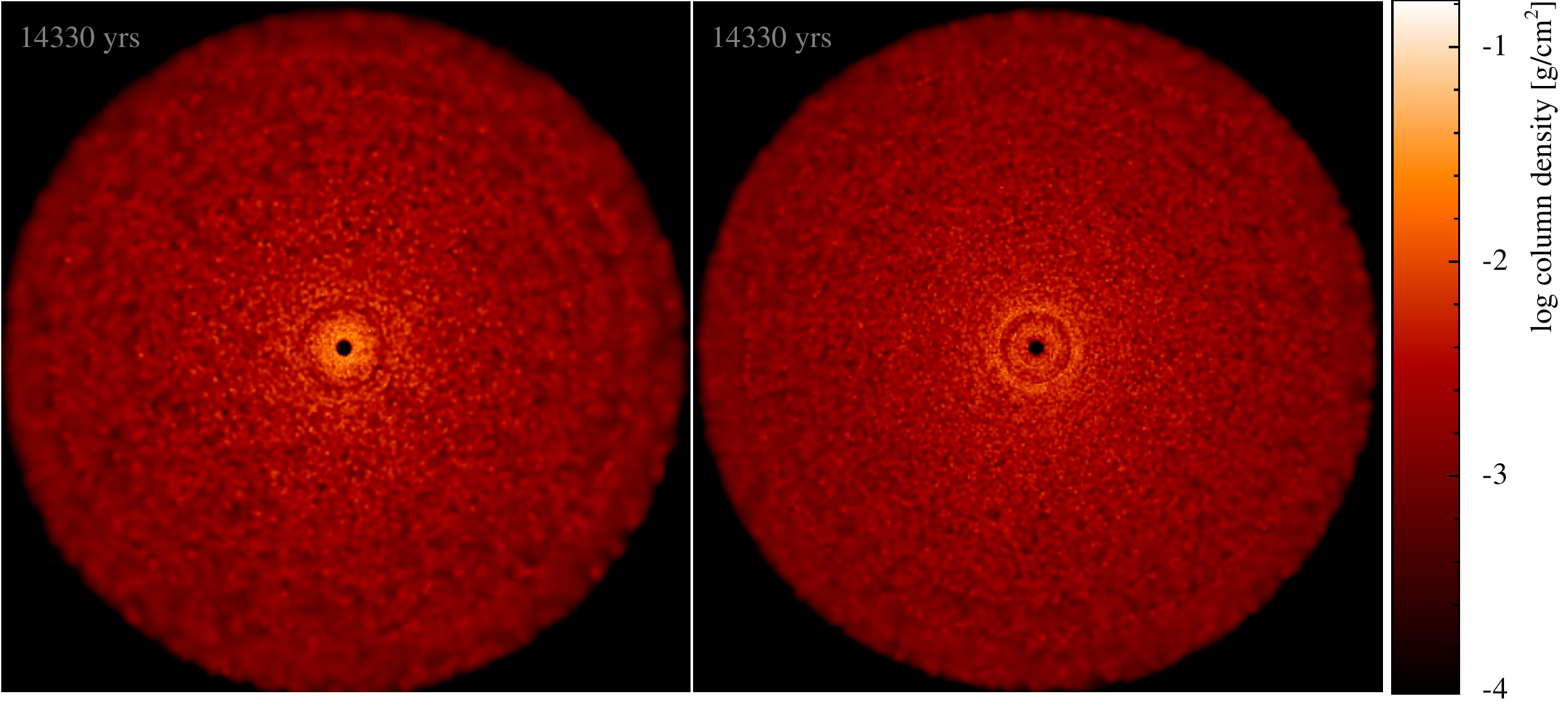} 
   \caption{Rendering of the density for the dust of a typical T-Tauri Star protoplanerary disc using $2\times10^{5}$ SPH particles, using an explicit time integration in the linear Epstein regime (left panel) and an implicit integrator in the full non-linear Epstein drag regimes (right panel).}
   \label{fig:compare}
\end{figure*}

Fig.~\ref{fig:compare} shows a face-on view of a protoplanetary disc, integrating the linear Epstein regime (left panel) and the full non-linear Epstein drag (right panel). The dust distributions are not found to exhibit significant discrepancies. In the non-linear drag regime case however, the dust distribution is slightly smoother since the drag (and thus, the coupling with the gas phase) is more efficient.

\begin{figure}
   \centering
   \includegraphics[angle=0, width=0.9\columnwidth]{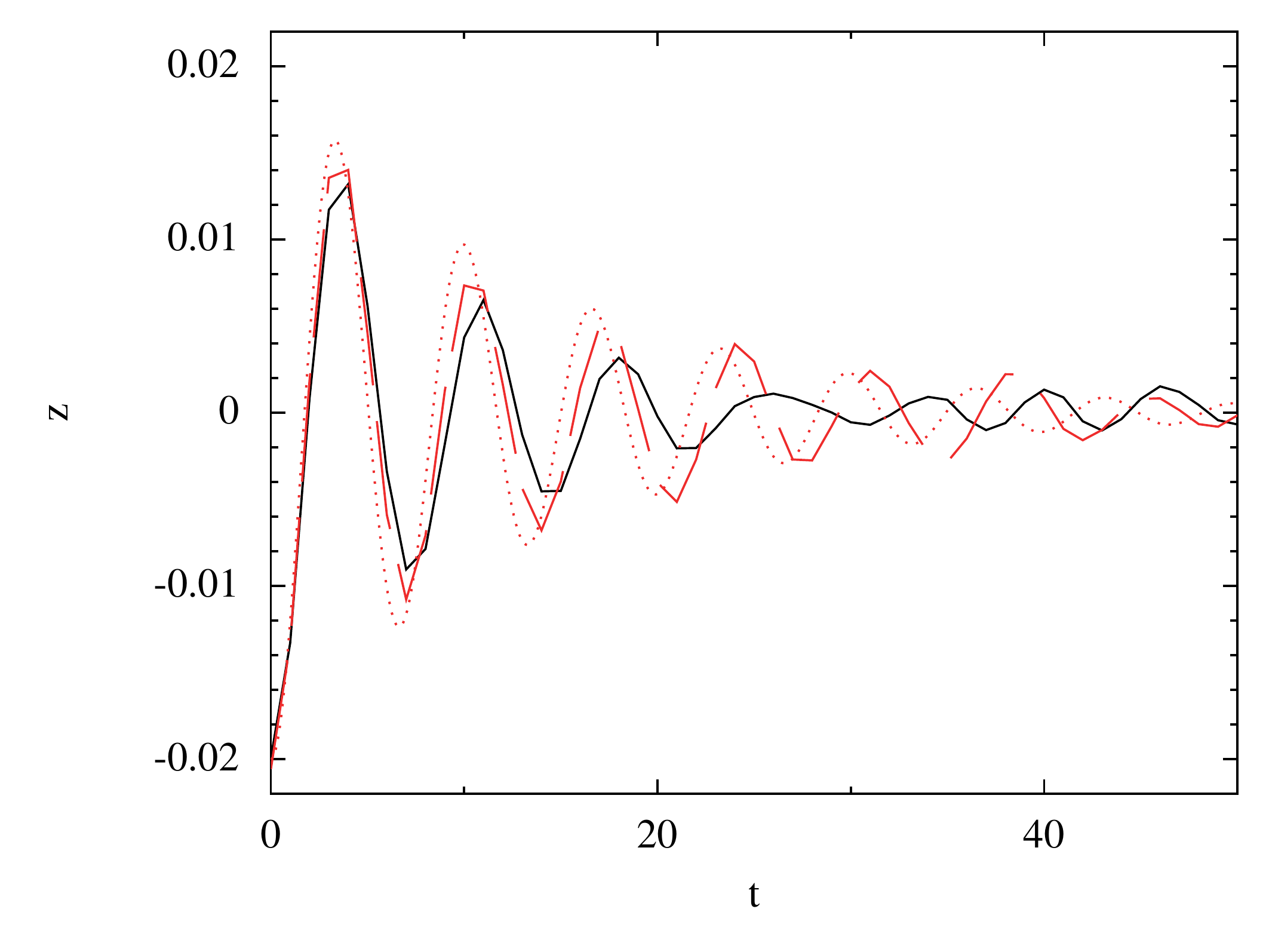} 
   \caption{Vertical settling of a dust grain ($1$cm in size) initially located at $r_{0}=100$ AU and $z_{0}=2$ AU (solid/black), integrating implicitly the non-linear Epstein regime. SPH results are compared to the explicit integration of the linear Epstein regime (dashed/red) and the estimation given by the damped harmonic oscillator approximation (pointed/red). In the full non-linear drag regime, the settling is more efficient than for the linear case since the vertical oscillations in the dust motion reaches a fraction $z_{0}/H$ of the sound speed.}
   \label{fig:settling}
\end{figure}

Fig.~\ref{fig:settling} compares the vertical motion of a dust grain initially located at $z=z_{0}$ using the linear (explicit integration) and the full non linear (implicit integration) Epstein regimes. In the full non-linear case, the settling is more efficient since the vertical differential velocity between the dust grains and the gas in the mid plane of the disc reaches a fraction $z_{0}/H$ of the sound speed, meaning that the non-linear terms are no longer negligible.

\section{Conclusions}

We have extended the SPH formalism for two-fluid dust and gas mixtures developed in Paper~I to handle the drag regimes usually encountered in a large range of astrophysical contexts. Specifically, our algorithm is now designed to treat the dynamics of grains surrounded by a dilute medium (Epstein regime) or dense fluid (Stokes regime), for which the drag force can be either linear or non-linear with respect to the differential velocity between the gas and the dust.

Particular attention has been paid to developing an implicit timestepping scheme to efficiently simulate the case of high drag, extending the scheme proposed by \citet{Monaghan1997} which we found to be unsatisfactory. We have presented a new pairwise implicit scheme that, like the \citet{Monaghan1997} scheme, preserves the exact conservation of linear and angular momentum but, unlike the \citet{Monaghan1997} scheme, i) provides control over the accuracy of the iterative procedure and ii) can incorporate non-linear terms for both Epstein and Stokes drag. We found that  when the ratio $r$ between the the timestep and the drag stopping time is $1 \lesssim r \lesssim 1000$, the implicit timestepping is faster than a standard explicit integration. However, at higher values of $r$, the algorithm is less efficient.

The accuracy of the generalised algorithm is benchmarked against the suite of test problems presented in Paper~I. In particular, the solutions obtained for the \textsc{dustybox} problem are compared to their known analytic solutions for a large range of non-linear drag regimes and the solutions of the \textsc{dustywave}, \textsc{dustyshock}, \textsc{dustysedov} and \textsc{dustydisc} problems are benchmarked against converged results obtained with explicit timestepping.

The two key issues addressed in this paper complete the study of our algorithm developed in Paper~I. Our intention is to apply it to various astrophysical problems involving gas and dust mixtures in star and planet formation. A first application is given in \citet{Ayliffe2011}.

\section*{Acknowledgments}

We thank Ben Ayliffe and Matthew Bate and Joe Monaghan for useful discussions and comments. Figures have been produced using \textsc{splash} \citep{splashpaper} with the new \textsc{giza} backend by DP and James Wetter. We are grateful to the Australian Research Council for funding via Discovery project grant DP1094585.

\bibliography{DustSPH}

\label{lastpage}
\end{document}

%% file: journaux.tex
%
%  These Macros are taken from the AAS TeX macro package version 4.0.
%  Include this file in your LaTeX source only if you are not using
%  the AAS TeX macro package and need to resolve the macro definitions
%  in the BibTeX entries returned by the ADS abstract service.
%
%  For more information on the AASTeX macro package, please see the URL
%	http://www.aas.org/publications/aastex.html
%  For more information about ADS abstract server, please see the URL
%	http://adswww.harvard.edu/ads_abstracts.html
%

% Abbreviations for journals.  The object here is to provide authors
% with convenient shorthands for the most "popular" (often-cited)
% journals; the author can use these markup tags without being concerned
% about the exact form of the journal abbreviation, or its formatting.
% It is up to the keeper of the macros to make sure the macros expand
% to the proper text.  If macro package writers agree to all use the
% same TeX command name, authors only have to remember one thing, and
% the style file will take care of editorial preferences.  This also
% applies when a single journal decides to revamp its abbreviating
% scheme, as happened with the ApJ (Abt 1991).

\def\jnl@style{\it}
%commente par Seb
\def\aaref@jnl#1{{\jnl@style#1}}
%ref remplace par aaref pour eviter conflit...

\def\aaref@jnl#1{{\jnl@style#1}}

\def\aj{\aaref@jnl{AJ}}                   % Astronomical Journal
\def\araa{\aaref@jnl{ARA\&A}}             % Annual Review of Astron and Astrophys
\def\apj{\aaref@jnl{ApJ}}                 % Astrophysical Journal
\def\apjl{\aaref@jnl{ApJ}}                % Astrophysical Journal, Letters
\def\apjs{\aaref@jnl{ApJS}}               % Astrophysical Journal, Supplement
\def\ao{\aaref@jnl{Appl.~Opt.}}           % Applied Optics
\def\apss{\aaref@jnl{Ap\&SS}}             % Astrophysics and Space Science
\def\aap{\aaref@jnl{A\&A}}                % Astronomy and Astrophysics
\def\aapr{\aaref@jnl{A\&A~Rev.}}          % Astronomy and Astrophysics Reviews
\def\aaps{\aaref@jnl{A\&AS}}              % Astronomy and Astrophysics, Supplement
\def\azh{\aaref@jnl{AZh}}                 % Astronomicheskii Zhurnal
\def\baas{\aaref@jnl{BAAS}}               % Bulletin of the AAS
\def\jrasc{\aaref@jnl{JRASC}}             % Journal of the RAS of Canada
\def\memras{\aaref@jnl{MmRAS}}            % Memoirs of the RAS
\def\mnras{\aaref@jnl{MNRAS}}             % Monthly Notices of the RAS
\def\pra{\aaref@jnl{Phys.~Rev.~A}}        % Physical Review A: General Physics
\def\prb{\aaref@jnl{Phys.~Rev.~B}}        % Physical Review B: Solid State
\def\prc{\aaref@jnl{Phys.~Rev.~C}}        % Physical Review C
\def\prd{\aaref@jnl{Phys.~Rev.~D}}        % Physical Review D
\def\pre{\aaref@jnl{Phys.~Rev.~E}}        % Physical Review E
\def\prl{\aaref@jnl{Phys.~Rev.~Lett.}}    % Physical Review Letters
\def\pasp{\aaref@jnl{PASP}}               % Publications of the ASP
\def\pasj{\aaref@jnl{PASJ}}               % Publications of the ASJ
\def\qjras{\aaref@jnl{QJRAS}}             % Quarterly Journal of the RAS
\def\skytel{\aaref@jnl{S\&T}}             % Sky and Telescope
\def\solphys{\aaref@jnl{Sol.~Phys.}}      % Solar Physics
\def\sovast{\aaref@jnl{Soviet~Ast.}}      % Soviet Astronomy
\def\ssr{\aaref@jnl{Space~Sci.~Rev.}}     % Space Science Reviews
\def\zap{\aaref@jnl{ZAp}}                 % Zeitschrift fuer Astrophysik
\def\nat{\aaref@jnl{Nature}}              % Nature
\def\iaucirc{\aaref@jnl{IAU~Circ.}}       % IAU Cirulars
\def\aplett{\aaref@jnl{Astrophys.~Lett.}} % Astrophysics Letters
\def\apspr{\aaref@jnl{Astrophys.~Space~Phys.~Res.}}
                % Astrophysics Space Physics Research
\def\bain{\aaref@jnl{Bull.~Astron.~Inst.~Netherlands}} 
                % Bulletin Astronomical Institute of the Netherlands
\def\fcp{\aaref@jnl{Fund.~Cosmic~Phys.}}  % Fundamental Cosmic Physics
\def\gca{\aaref@jnl{Geochim.~Cosmochim.~Acta}}   % Geochimica Cosmochimica Acta
\def\grl{\aaref@jnl{Geophys.~Res.~Lett.}} % Geophysics Research Letters
\def\jcp{\aaref@jnl{J.~Chem.~Phys.}}      % Journal of Chemical Physics
\def\jgr{\aaref@jnl{J.~Geophys.~Res.}}    % Journal of Geophysics Research
\def\jqsrt{\aaref@jnl{J.~Quant.~Spec.~Radiat.~Transf.}}
                % Journal of Quantitiative Spectroscopy and Radiative Transfer
\def\memsai{\aaref@jnl{Mem.~Soc.~Astron.~Italiana}}
                % Mem. Societa Astronomica Italiana
\def\nphysa{\aaref@jnl{Nucl.~Phys.~A}}   % Nuclear Physics A
\def\physrep{\aaref@jnl{Phys.~Rep.}}   % Physics Reports
\def\physscr{\aaref@jnl{Phys.~Scr}}   % Physica Scripta
\def\planss{\aaref@jnl{Planet.~Space~Sci.}}   % Planetary Space Science
\def\procspie{\aaref@jnl{Proc.~SPIE}}   % Proceedings of the SPIE

\let\astap=\aap
\let\apjlett=\apjl
\let\apjsupp=\apjs
\let\applopt=\ao